\documentstyle[eqsecnum,aps,preprint]{revtex}
\begin{document}
\draft
\def\PsfigVersion{1.9}
\ifx\undefined\psfig\else \fi

%

\let\LaTeXAtSign=\@
\let\@=\relax
\edef\psfigRestoreAt{\catcode`\@=\number\catcode`@\relax}
\catcode`\@=11\relax
\newwrite\@unused
\def\ps@typeout#1{{\let\protect\string\immediate\write\@unused{#1}}}
\ps@typeout{psfig/tex \PsfigVersion}


\def\figurepath{./}
\def\psfigurepath#1{\edef\figurepath{#1}}

%
%
\def\@nnil{\@nil}
\def\@empty{}
\def\@psdonoop#1\@@#2#3{}
\def\@psdo#1:=#2\do#3{\edef\@psdotmp{#2}\ifx\@psdotmp\@empty \else
    \expandafter\@psdoloop#2,\@nil,\@nil\@@#1{#3}\fi}
\def\@psdoloop#1,#2,#3\@@#4#5{\def#4{#1}\ifx #4\@nnil \else
       #5\def#4{#2}\ifx #4\@nnil \else#5\@ipsdoloop #3\@@#4{#5}\fi\fi}
\def\@ipsdoloop#1,#2\@@#3#4{\def#3{#1}\ifx #3\@nnil 
       \let\@nextwhile=\@psdonoop \else
      #4\relax\let\@nextwhile=\@ipsdoloop\fi\@nextwhile#2\@@#3{#4}}
\def\@tpsdo#1:=#2\do#3{\xdef\@psdotmp{#2}\ifx\@psdotmp\@empty \else
    \@tpsdoloop#2\@nil\@nil\@@#1{#3}\fi}
\def\@tpsdoloop#1#2\@@#3#4{\def#3{#1}\ifx #3\@nnil 
       \let\@nextwhile=\@psdonoop \else
      #4\relax\let\@nextwhile=\@tpsdoloop\fi\@nextwhile#2\@@#3{#4}}
%
\ifx\undefined\fbox
\newdimen\fboxrule
\newdimen\fboxsep
\newdimen\ps@tempdima
\newbox\ps@tempboxa
\fboxsep = 3pt
\fboxrule = .4pt
\long\def\fbox#1{\leavevmode\setbox\ps@tempboxa\hbox{#1}\ps@tempdima\fboxrule
    \advance\ps@tempdima \fboxsep \advance\ps@tempdima \dp\ps@tempboxa
   \hbox{\lower \ps@tempdima\hbox
  {\vbox{\hrule height \fboxrule
          \hbox{\vrule width \fboxrule \hskip\fboxsep
          \vbox{\vskip\fboxsep \box\ps@tempboxa\vskip\fboxsep}\hskip 
                 \fboxsep\vrule width \fboxrule}
                 \hrule height \fboxrule}}}}
\fi
%
%
\newread\ps@stream
\newif\ifnot@eof       
\newif\if@noisy        
\newif\if@atend        
\newif\if@psfile       
%
%
{\catcode`\%=12\global\gdef\epsf@start{
\def\epsf@PS{PS}
\def\epsf@getbb#1{%
%
%
\openin\ps@stream=#1
\ifeof\ps@stream\ps@typeout{Error, File #1 not found}\else
%
%
   {\not@eoftrue \chardef\other=12
    \def\do##1{\catcode`##1=\other}\dospecials \catcode`\ =10
    \loop
       \if@psfile
	  \read\ps@stream to \epsf@fileline
       \else{
	  \obeyspaces
          \read\ps@stream to \epsf@tmp\global\let\epsf@fileline\epsf@tmp}
       \fi
       \ifeof\ps@stream\not@eoffalse\else
%
%
       \if@psfile\else
       \expandafter\epsf@test\epsf@fileline:. \\%
       \fi
%
%
          \expandafter\epsf@aux\epsf@fileline:. \\%
       \fi
   \ifnot@eof\repeat
   }\closein\ps@stream\fi}%
%
%
\long\def\epsf@test#1#2#3:#4\\{\def\epsf@testit{#1#2}
			\ifx\epsf@testit\epsf@start\else
\ps@typeout{Warning! File does not start with `\epsf@start'.  It may not be a PostScript file.}
			\fi
			\@psfiletrue} 
%
%
{\catcode`\%=12\global\let\epsf@percent=
%
%
%
\long\def\epsf@aux#1#2:#3\\{\ifx#1\epsf@percent
   \def\epsf@testit{#2}\ifx\epsf@testit\epsf@bblit
	\@atendfalse
        \epsf@atend #3 . \\%
	\if@atend	
	   \if@verbose{
		\ps@typeout{psfig: found `(atend)'; continuing search}
	   }\fi
        \else
        \epsf@grab #3 . . . \\%
        \not@eoffalse
        \global\no@bbfalse
        \fi
   \fi\fi}%
%
%
\def\epsf@grab #1 #2 #3 #4 #5\\{%
   \global\def\epsf@llx{#1}\ifx\epsf@llx\empty
      \epsf@grab #2 #3 #4 #5 .\\\else
   \global\def\epsf@lly{#2}%
   \global\def\epsf@urx{#3}\global\def\epsf@ury{#4}\fi}%
%
%
\def\epsf@atendlit{(atend)} 
\def\epsf@atend #1 #2 #3\\{%
   \def\epsf@tmp{#1}\ifx\epsf@tmp\empty
      \epsf@atend #2 #3 .\\\else
   \ifx\epsf@tmp\epsf@atendlit\@atendtrue\fi\fi}


\chardef\psletter = 11 
\chardef\other = 12

\newif \ifdebug 
\newif\ifc@mpute 
\c@mputetrue 

\let\then = \relax
\def\r@dian{pt }
\let\r@dians = \r@dian
\let\dimensionless@nit = \r@dian
\let\dimensionless@nits = \dimensionless@nit
\def\internal@nit{sp }
\let\internal@nits = \internal@nit
\newif\ifstillc@nverging
\def \Mess@ge #1{\ifdebug \then \message {#1} \fi}

{ 
	\catcode `\@ = \psletter
	\gdef \nodimen {\expandafter \n@dimen \the \dimen}
	\gdef \term #1 #2 #3%
	       {\edef \t@ {\the #1}
		\edef \t@@ {\expandafter \n@dimen \the #2\r@dian}%
		\t@rm {\t@} {\t@@} {#3}%
	       }
	\gdef \t@rm #1 #2 #3%
	       {{%
		\count 0 = 0
		\dimen 0 = 1 \dimensionless@nit
		\dimen 2 = #2\relax
		\Mess@ge {Calculating term #1 of \nodimen 2}%
		\loop
		\ifnum	\count 0 < #1
		\then	\advance \count 0 by 1
			\Mess@ge {Iteration \the \count 0 \space}%
			\Multiply \dimen 0 by {\dimen 2}%
			\Mess@ge {After multiplication, term = \nodimen 0}%
			\Divide \dimen 0 by {\count 0}%
			\Mess@ge {After division, term = \nodimen 0}%
		\repeat
		\Mess@ge {Final value for term #1 of 
				\nodimen 2 \space is \nodimen 0}%
		\xdef \Term {#3 = \nodimen 0 \r@dians}%
		\aftergroup \Term
	       }}
	\catcode `\p = \other
	\catcode `\t = \other
	\gdef \n@dimen #1pt{#1} 
}

\def \Divide #1by #2{\divide #1 by #2} 

\def \Multiply #1by #2
       {{
	\count 0 = #1\relax
	\count 2 = #2\relax
	\count 4 = 65536
	\Mess@ge {Before scaling, count 0 = \the \count 0 \space and
			count 2 = \the \count 2}%
	\ifnum	\count 0 > 32767 
	\then	\divide \count 0 by 4
		\divide \count 4 by 4
	\else	\ifnum	\count 0 < -32767
		\then	\divide \count 0 by 4
			\divide \count 4 by 4
		\else
		\fi
	\fi
	\ifnum	\count 2 > 32767 
	\then	\divide \count 2 by 4
		\divide \count 4 by 4
	\else	\ifnum	\count 2 < -32767
		\then	\divide \count 2 by 4
			\divide \count 4 by 4
		\else
		\fi
	\fi
	\multiply \count 0 by \count 2
	\divide \count 0 by \count 4
	\xdef \product {#1 = \the \count 0 \internal@nits}%
	\aftergroup \product
       }}

\def\r@duce{\ifdim\dimen0 > 90\r@dian \then   
		\multiply\dimen0 by -1
		\advance\dimen0 by 180\r@dian
		\r@duce
	    \else \ifdim\dimen0 < -90\r@dian \then  
		\advance\dimen0 by 360\r@dian
		\r@duce
		\fi
	    \fi}

\def\Sine#1%
       {{%
	\dimen 0 = #1 \r@dian
	\r@duce
	\ifdim\dimen0 = -90\r@dian \then
	   \dimen4 = -1\r@dian
	   \c@mputefalse
	\fi
	\ifdim\dimen0 = 90\r@dian \then
	   \dimen4 = 1\r@dian
	   \c@mputefalse
	\fi
	\ifdim\dimen0 = 0\r@dian \then
	   \dimen4 = 0\r@dian
	   \c@mputefalse
	\fi
	\ifc@mpute \then
		\divide\dimen0 by 180
		\dimen0=3.141592654\dimen0
		\dimen 2 = 3.1415926535897963\r@dian 
		\divide\dimen 2 by 2 
		\Mess@ge {Sin: calculating Sin of \nodimen 0}%
		\count 0 = 1 
		\dimen 2 = 1 \r@dian 
		\dimen 4 = 0 \r@dian 
		\loop
			\ifnum	\dimen 2 = 0 
			\then	\stillc@nvergingfalse 
			\else	\stillc@nvergingtrue
			\fi
			\ifstillc@nverging 
			\then	\term {\count 0} {\dimen 0} {\dimen 2}%
				\advance \count 0 by 2
				\count 2 = \count 0
				\divide \count 2 by 2
				\ifodd	\count 2 
				\then	\advance \dimen 4 by \dimen 2
				\else	\advance \dimen 4 by -\dimen 2
				\fi
		\repeat
	\fi		
			\xdef \sine {\nodimen 4}%
       }}

\def\Cosine#1{\ifx\sine\UnDefined\edef\Savesine{\relax}\else
		             \edef\Savesine{\sine}\fi
	{\dimen0=#1\r@dian\advance\dimen0 by 90\r@dian
	 \Sine{\nodimen 0}
	 \xdef\cosine{\sine}
	 \xdef\sine{\Savesine}}}	      

\def\psdraft{
	\def\@psdraft{0}
}
\def\psfull{
	\def\@psdraft{100}
}

\psfull

\newif\if@scalefirst
\def\psscalefirst{\@scalefirsttrue}
\def\psrotatefirst{\@scalefirstfalse}
\psrotatefirst

\newif\if@draftbox
\def\psnodraftbox{
	\@draftboxfalse
}
\def\psdraftbox{
	\@draftboxtrue
}
\@draftboxtrue

\newif\if@prologfile
\newif\if@postlogfile
\def\pssilent{
	\@noisyfalse
}
\def\psnoisy{
	\@noisytrue
}
\psnoisy
\newif\if@bbllx
\newif\if@bblly
\newif\if@bburx
\newif\if@bbury
\newif\if@height
\newif\if@width
\newif\if@rheight
\newif\if@rwidth
\newif\if@angle
\newif\if@clip
\newif\if@verbose
\def\@p@@sclip#1{\@cliptrue}

\newif\if@decmpr


\def\@p@@sfigure#1{\def\@p@sfile{null}\def\@p@sbbfile{null}
	        \openin1=#1.bb
		\ifeof1\closein1
	        	\openin1=\figurepath#1.bb
			\ifeof1\closein1
			        \openin1=#1
				\ifeof1\closein1%
				       \openin1=\figurepath#1
					\ifeof1
					   \ps@typeout{Error, File #1 not found}
						\if@bbllx\if@bblly
				   		\if@bburx\if@bbury
			      				\def\@p@sfile{#1}%
			      				\def\@p@sbbfile{#1}%
							\@decmprfalse
				  	   	\fi\fi\fi\fi
					\else\closein1
				    		\def\@p@sfile{\figurepath#1}%
				    		\def\@p@sbbfile{\figurepath#1}%
						\@decmprfalse
	                       		\fi%
			 	\else\closein1%
					\def\@p@sfile{#1}
					\def\@p@sbbfile{#1}
					\@decmprfalse
			 	\fi
			\else
				\def\@p@sfile{\figurepath#1}
				\def\@p@sbbfile{\figurepath#1.bb}
				\@decmprtrue
			\fi
		\else
			\def\@p@sfile{#1}
			\def\@p@sbbfile{#1.bb}
			\@decmprtrue
		\fi}

\def\@p@@sfile#1{\@p@@sfigure{#1}}

\def\@p@@sbbllx#1{
		\@bbllxtrue
		\dimen100=#1
		\edef\@p@sbbllx{\number\dimen100}
}
\def\@p@@sbblly#1{
		\@bbllytrue
		\dimen100=#1
		\edef\@p@sbblly{\number\dimen100}
}
\def\@p@@sbburx#1{
		\@bburxtrue
		\dimen100=#1
		\edef\@p@sbburx{\number\dimen100}
}
\def\@p@@sbbury#1{
		\@bburytrue
		\dimen100=#1
		\edef\@p@sbbury{\number\dimen100}
}
\def\@p@@sheight#1{
		\@heighttrue
		\dimen100=#1
   		\edef\@p@sheight{\number\dimen100}
}
\def\@p@@swidth#1{
		\@widthtrue
		\dimen100=#1
		\edef\@p@swidth{\number\dimen100}
}
\def\@p@@srheight#1{
		\@rheighttrue
		\dimen100=#1
		\edef\@p@srheight{\number\dimen100}
}
\def\@p@@srwidth#1{
		\@rwidthtrue
		\dimen100=#1
		\edef\@p@srwidth{\number\dimen100}
}
\def\@p@@sangle#1{
		\@angletrue
		\edef\@p@sangle{#1} 
}
\def\@p@@ssilent#1{ 
		\@verbosefalse
}
\def\@p@@sprolog#1{\@prologfiletrue\def\@prologfileval{#1}}
\def\@p@@spostlog#1{\@postlogfiletrue\def\@postlogfileval{#1}}
\def\@cs@name#1{\csname #1\endcsname}
\def\@setparms#1=#2,{\@cs@name{@p@@s#1}{#2}}
%
%
\def\ps@init@parms{
		\@bbllxfalse \@bbllyfalse
		\@bburxfalse \@bburyfalse
		\@heightfalse \@widthfalse
		\@rheightfalse \@rwidthfalse
		\def\@p@sbbllx{}\def\@p@sbblly{}
		\def\@p@sbburx{}\def\@p@sbbury{}
		\def\@p@sheight{}\def\@p@swidth{}
		\def\@p@srheight{}\def\@p@srwidth{}
		\def\@p@sangle{0}
		\def\@p@sfile{} \def\@p@sbbfile{}
		\def\@p@scost{10}
		\def\@sc{}
		\@prologfilefalse
		\@postlogfilefalse
		\@clipfalse
		\if@noisy
			\@verbosetrue
		\else
			\@verbosefalse
		\fi
}
%
%
\def\parse@ps@parms#1{
	 	\@psdo\@psfiga:=#1\do
		   {\expandafter\@setparms\@psfiga,}}
%
%
\newif\ifno@bb
\def\bb@missing{
	\if@verbose{
		\ps@typeout{psfig: searching \@p@sbbfile \space  for bounding box}
	}\fi
	\no@bbtrue
	\epsf@getbb{\@p@sbbfile}
        \ifno@bb \else \bb@cull\epsf@llx\epsf@lly\epsf@urx\epsf@ury\fi
}	
\def\bb@cull#1#2#3#4{
	\dimen100=#1 bp\edef\@p@sbbllx{\number\dimen100}
	\dimen100=#2 bp\edef\@p@sbblly{\number\dimen100}
	\dimen100=#3 bp\edef\@p@sbburx{\number\dimen100}
	\dimen100=#4 bp\edef\@p@sbbury{\number\dimen100}
	\no@bbfalse
}
\newdimen\p@intvaluex
\newdimen\p@intvaluey
\def\rotate@#1#2{{\dimen0=#1 sp\dimen1=#2 sp
		  \global\p@intvaluex=\cosine\dimen0
		  \dimen3=\sine\dimen1
		  \global\advance\p@intvaluex by -\dimen3
		  \global\p@intvaluey=\sine\dimen0
		  \dimen3=\cosine\dimen1
		  \global\advance\p@intvaluey by \dimen3
		  }}
\def\compute@bb{
		\no@bbfalse
		\if@bbllx \else \no@bbtrue \fi
		\if@bblly \else \no@bbtrue \fi
		\if@bburx \else \no@bbtrue \fi
		\if@bbury \else \no@bbtrue \fi
		\ifno@bb \bb@missing \fi
		\ifno@bb \ps@typeout{FATAL ERROR: no bb supplied or found}
			\no-bb-error
		\fi
		%
%
		\count203=\@p@sbburx
		\count204=\@p@sbbury
		\advance\count203 by -\@p@sbbllx
		\advance\count204 by -\@p@sbblly
		\edef\ps@bbw{\number\count203}
		\edef\ps@bbh{\number\count204}
		\if@angle 
			\Sine{\@p@sangle}\Cosine{\@p@sangle}
	        	{\dimen100=\maxdimen\xdef\r@p@sbbllx{\number\dimen100}
					    \xdef\r@p@sbblly{\number\dimen100}
			                    \xdef\r@p@sbburx{-\number\dimen100}
					    \xdef\r@p@sbbury{-\number\dimen100}}
%
                        \def\minmaxtest{
			   \ifnum\number\p@intvaluex<\r@p@sbbllx
			      \xdef\r@p@sbbllx{\number\p@intvaluex}\fi
			   \ifnum\number\p@intvaluex>\r@p@sbburx
			      \xdef\r@p@sbburx{\number\p@intvaluex}\fi
			   \ifnum\number\p@intvaluey<\r@p@sbblly
			      \xdef\r@p@sbblly{\number\p@intvaluey}\fi
			   \ifnum\number\p@intvaluey>\r@p@sbbury
			      \xdef\r@p@sbbury{\number\p@intvaluey}\fi
			   }
			\rotate@{\@p@sbbllx}{\@p@sbblly}
			\minmaxtest
			\rotate@{\@p@sbbllx}{\@p@sbbury}
			\minmaxtest
			\rotate@{\@p@sbburx}{\@p@sbblly}
			\minmaxtest
			\rotate@{\@p@sbburx}{\@p@sbbury}
			\minmaxtest
			\edef\@p@sbbllx{\r@p@sbbllx}\edef\@p@sbblly{\r@p@sbblly}
			\edef\@p@sbburx{\r@p@sbburx}\edef\@p@sbbury{\r@p@sbbury}
		\fi
		\count203=\@p@sbburx
		\count204=\@p@sbbury
		\advance\count203 by -\@p@sbbllx
		\advance\count204 by -\@p@sbblly
		\edef\@bbw{\number\count203}
		\edef\@bbh{\number\count204}
}
%
%
\def\in@hundreds#1#2#3{\count240=#2 \count241=#3
		     \count100=\count240	
		     \divide\count100 by \count241
		     \count101=\count100
		     \multiply\count101 by \count241
		     \advance\count240 by -\count101
		     \multiply\count240 by 10
		     \count101=\count240	
		     \divide\count101 by \count241
		     \count102=\count101
		     \multiply\count102 by \count241
		     \advance\count240 by -\count102
		     \multiply\count240 by 10
		     \count102=\count240	
		     \divide\count102 by \count241
		     \count200=#1\count205=0
		     \count201=\count200
			\multiply\count201 by \count100
		 	\advance\count205 by \count201
		     \count201=\count200
			\divide\count201 by 10
			\multiply\count201 by \count101
			\advance\count205 by \count201
		     \count201=\count200
			\divide\count201 by 100
			\multiply\count201 by \count102
			\advance\count205 by \count201
		     \edef\@result{\number\count205}
}
\def\compute@wfromh{
		\in@hundreds{\@p@sheight}{\@bbw}{\@bbh}
		\edef\@p@swidth{\@result}
}
\def\compute@hfromw{
	        \in@hundreds{\@p@swidth}{\@bbh}{\@bbw}
		\edef\@p@sheight{\@result}
}
\def\compute@handw{
		\if@height 
			\if@width
			\else
				\compute@wfromh
			\fi
		\else 
			\if@width
				\compute@hfromw
			\else
				\edef\@p@sheight{\@bbh}
				\edef\@p@swidth{\@bbw}
			\fi
		\fi
}
\def\compute@resv{
		\if@rheight \else \edef\@p@srheight{\@p@sheight} \fi
		\if@rwidth \else \edef\@p@srwidth{\@p@swidth} \fi
}
%
\def\compute@sizes{
	\compute@bb
	\if@scalefirst\if@angle
	\if@width
	   \in@hundreds{\@p@swidth}{\@bbw}{\ps@bbw}
	   \edef\@p@swidth{\@result}
	\fi
	\if@height
	   \in@hundreds{\@p@sheight}{\@bbh}{\ps@bbh}
	   \edef\@p@sheight{\@result}
	\fi
	\fi\fi
	\compute@handw
	\compute@resv}

%
%
\def\psfig#1{\vbox {
	%
	\ps@init@parms
	\parse@ps@parms{#1}
	\compute@sizes
	\ifnum\@p@scost<\@psdraft{
		\special{ps::[begin] 	\@p@swidth \space \@p@sheight \space
				\@p@sbbllx \space \@p@sbblly \space
				\@p@sbburx \space \@p@sbbury \space
				startTexFig \space }
		\if@angle
			\special {ps:: \@p@sangle \space rotate \space} 
		\fi
		\if@clip{
			\if@verbose{
				\ps@typeout{(clip)}
			}\fi
			\special{ps:: doclip \space }
		}\fi
		\if@prologfile
		    \special{ps: plotfile \@prologfileval \space } \fi
		\if@decmpr{
			\if@verbose{
				\ps@typeout{psfig: including \@p@sfile.Z \space }
			}\fi
			\special{ps: plotfile "`zcat \@p@sfile.Z" \space }
		}\else{
			\if@verbose{
				\ps@typeout{psfig: including \@p@sfile \space }
			}\fi
			\special{ps: plotfile \@p@sfile \space }
		}\fi
		\if@postlogfile
		    \special{ps: plotfile \@postlogfileval \space } \fi
		\special{ps::[end] endTexFig \space }
		\vbox to \@p@srheight sp{
			\hbox to \@p@srwidth sp{
				\hss
			}
		\vss
		}
	}\else{
		\if@draftbox{		
			\hbox{\frame{\vbox to \@p@srheight sp{
			\vss
			\hbox to \@p@srwidth sp{ \hss \@p@sfile \hss }
			\vss
			}}}
		}\else{
			\vbox to \@p@srheight sp{
			\vss
			\hbox to \@p@srwidth sp{\hss}
			\vss
			}
		}\fi

	}\fi
}}
\psfigRestoreAt
\let\@=\LaTeXAtSign

\title{Forest fires and other examples of self-organized criticality}
\author{Siegfried Clar, Barbara Drossel, and Franz Schwabl}
\address{Institut f\"ur Theoretische Physik, \\
         Physik--Department der Technischen Universit\"at M\"unchen, \\
         James--Franck--Str., D--85747 Garching, Germany}
\date{\today}
\maketitle
\begin{abstract}
We review the properties of the self--organized critical (SOC) forest--fire
model. The paradigm of self--organized criticality refers to the tendency of
certain large dissipative systems to drive themselves into a critical state
independent of the initial conditions and without fine-tuning of the 
parameters. 
After an introduction, we define the rules of the model 
and discuss various large--scale structures which may appear in this
system. The origin of the critical behavior is explained, critical exponents
are introduced,
and scaling relations between the exponents
are derived. Results of computer simulations
and analytical calculations are summarized. The existence of an upper
critical dimension and the universality of the critical behavior under
changes of lattice symmetry or the introduction of immunity are discussed. 
A survey of interesting modifications of the forest--fire model is
given.
Finally, several other important SOC models are briefly described.
\end{abstract}

\section{Introduction}
\label{u1}

Driven dissipative systems far from thermodynamic equilibrium show a rich 
variety of patterns. Since they receive a permanent input of energy, they 
can maintain states which are highly ordered or complex. Ubiquitous examples 
for complex structures are {\em fractals} \cite{man83,bun91} as well as their 
temporal counterpart, {\em $1/f$--noise} \cite{pre78,dut81}. Fractals
are self-similar structures that look the same on different scales
of observation, since they have no intrinsic length scale. Their spatial
correlation functions are power laws.
Well-known examples are mountain landscapes and coastlines. 
$1/f$--noise is the temporal equivalent of fractals.
Its name indicates that the Fourier transform of the temporal 
correlation function is a power law of the form
$1/f^\alpha$ with $\alpha \approx 1$.
Like fractals, $1/f$--noise can be found in many natural systems, e.g.
undersea ocean currents or stock market prices. 

In 1987, Bak, Tang, and Wiesenfeld introduced the {\em sandpile model} 
which evolves into a critical state irrespective of initial conditions and
without fine tuning of para\-meters \cite{bak87,bak88}. Such extended
non-equilibrium systems are 
called {\em self--organized critical} (SOC) and exhibit power--law 
correlations in space
and time. The concept of SOC (for a review see \cite{bak94}) 
has attracted much
interest since it might explain part of the abundance of fractals and 
$1/f$--noise in nature and create a link between the two. 
Models for earthquakes 
\cite{ola92,chr92}, the evolution
of populations \cite{bak89,sne93}, the formation of clouds \cite{nag92} and
river networks \cite{tak92}, and many more have been introduced and 
investigated mainly by
computer simulations, but with few exceptions,
most of these SOC models still are barely understood. 

This review is mainly devoted to the {\em forest--fire model} (FFM) \cite{dro92}, which is a particularly simple example for open systems which shows nevertheless a variety of different structures, depending on the parameters. In the limit of a double separation of time scales, the FFM becomes SOC, with a power--law distribution of fires and forest clusters. 

The FFM is a stochastic cellular automaton which is defined 
on a $d$--dimensional hyper-cubic lattice with $L^d$ sites. Each site
is occupied by
a tree, a burning tree, or it is empty. During one time step, the system is
parallely updated according to the following rules
\begin{enumerate}
\item Burning tree $\longrightarrow$ empty site.
\item Tree $\longrightarrow$ burning tree with probability $1-g$ if at least 
one nearest neighbor is burning. 
\item Tree $\longrightarrow $ burning tree with probability $f$ if no nearest 
neighbor is burning.
\item Empty site $\longrightarrow$ tree with probability $p$.
\end{enumerate}
An important application of the FFM comes from its close relation to
excitable media \cite{dro93a}, which comprise phenomena so different
as spreading of diseases, oscillating chemical reactions, 
propagation of electrical activity in neurons or heart muscles,
spiral galaxies, and many more 
(For a review on excitable systems see e.g. \cite{tys88,mer92}).
These systems essentially have three states which are called 
quiescent ($\corresponds$ tree), excited ($\corresponds$ burning tree),
and refractory ($\corresponds$ empty site). Excitation spreads
from one place to its neighbors if they are quiescent. After excitation,
a refractory site needs some time to recover its quiescent state. Since
excitable systems are often deterministic, it is, however, not always possible
to describe their behavior in the framework of the FFM, where some of the 
parameters are stochastic.

Turning back to the FFM, we consider a system with arbitrary initial 
conditions. After a transition period the system approaches 
a steady state the
properties of which depend only on the parameter values. 
If the system size $L$ is large enough 
the properties of the steady state are also independent of the
boundary conditions.
Let $\rho_e$,
$\rho_t$, and $\rho_f$ be the mean density of empty sites, of trees,
and of burning trees in the steady state. These densities are
related by the equations
\begin{equation}
\rho_e+\rho_t+\rho_f=1 \label{luisa}
\end{equation}
and
\begin{equation}
\rho_f=p\rho_e. \label{conchita}
\end{equation}
The second relation says that the mean number of growing trees
equals the mean number of burning trees in the steady state. 
Depending on the magnitude of the parameters, a variety of large scale 
structures arise.

\subsection{Fire fronts and spirals}

In the limit of slow tree growth $p$ and without lightning and immunity,
fire only spreads from burning trees to their neighbors but does
not occur spontaneously.
In this version, which was originally introduced by P. Bak, K. Chen, and C.
Tang \cite{bak90}, the fire fronts become more and more regular
and spiral--shaped with decreasing $p$ \cite{gra91}.
A snapshot of this state is shown in Fig.~\ref{fig1}.
A more thorough analysis of the spatial and temporal correlations
in the system revealed the existence of characteristic length and
time scales proportional to $1/p$, both of which become increasingly
distinct with decreasing $p$ \cite{mos92}, indicating increasing determinism in this limit.
Spiral waves are very familiar in excitable media and have been 
observed in oscillating chemical reactions, the heart muscle, and 
the chemical signals of the amoeba dictyostelium discoideum
\cite{tys88,mer92}. It is also known that certain epidemics occur periodically
\cite{tys88}. They occur already in simple deterministic versions of the FFM \cite{gre78}.

\subsection{Percolation due to immunity}

When the immunity is nonzero, the fire fronts present for $g=0$ become more and more fuzzy with 
increasing immunity $g$, and the forest becomes denser \cite{dro93a}.
At a critical 
value $g = g_C(p)$ the fire density becomes zero and the forest density
becomes one. Figs.~\ref{fig2} and \ref{fig3} show 
two snapshots of the system for
values of $g$ far below $g_C(p)$ and near $g_C(p)$. At $g_C(p)$, the fire just
percolates through the system. Since sites are not permanently immune, this
kind of percolation is different from usual site percolation and is in the same universality class as directed percolation in $d + 1$ dimensions (the preferred direction corresponds to the time) \cite{alb95}. The critical
immunity  $g_C(p)$ increases with increasing $p$, since the fire then 
can return sooner to sites where it already has been.
Above $g_C(p)$ the 
steady state of the system is a completely dense forest. 
A similar model, using the language of spreading diseases has been studied independently in \cite{joh94}.

\subsection{Self-organized critical (SOC) behavior}

The SOC behavior occurs when the lightning probability is nonzero. For simplicity, we set the immunity to zero, but we will show later that the SOC state persists for $g > 0$. The ratio $p/f$ is a measure for the number of trees growing between two lightning strokes and therefore for the mean number of trees destroyed per lightning stroke. In the limit 
\begin{equation}
f \ll p\, , \label{luciana}
\end{equation}
there exist consequently large forest clusters and correlations over large distances. The model is SOC when tree growth is so slow that fire burns down even large clusters before new trees grow at their edge. This condition, which assures invariance under a change of the length scale, reads
\begin{equation}
p \ll T^{-1}(s_{\text{max}})\, , \label{maria}
\end{equation}
where $T(s_{\text{max}})$ is the time the fire needs to burn down a large
forest cluster. It diverges in the limit $f/p \to 0$ according to $T(s_{\text{max}})
\propto (f/p)^{-\nu'}$, with an appropriate exponent $\nu'$. 

In this case, the
dynamics of the system depend only on the ratio $f/p$, but not on $f$ and $p$
separately. When $f$ and $p$ are both decreased by the same factor, the overall
time scale of the system is also changed by this factor, but not the number of
trees that grow between two lightnings and therefore not the size distribution
of forest clusters and of fires.  
In the simulations, the condition (\ref{maria}) is most
easily realized by burning forest clusters instantaneously,
i.e. during one time step. This extreme limit of the SOC forest--fire
model has been invented independently by C. Henley \cite{hen89}.

The inequalities (\ref{luciana}) and (\ref{maria}) represent a 
{\em double separation of time scales}
\begin{equation}
(f/p)^{-\nu'} \ll p^{-1} \ll f^{-1}, 
\end{equation}
which is the condition for SOC behavior in
the FFM. The time in which a forest cluster burns down is much
shorter than the time in which a tree grows, which again is much shorter than
the time between two lightning occurrences. Separation of time scales is quite
frequent in nature, while the tuning of parameters to a certain finite value
only takes place accidentally. Thus, the FFM is critical over a
wide range of parameter values. A snapshot of the critical state is shown in 
Fig.~\ref{fig4}.

In the following, we will mainly discuss the properties of the SOC FFM, and we
will compare this model to other SOC systems, as there are e.g. 
the sandpile model,
the earthquake model, etc.
The outline of the remainder of this paper is as follows: 
In Sec.~\ref{u21}, the critical exponents of the forest-fire model
are defined, and scaling relations between them are derived.
Sec.~\ref{u22} presents the results of computer simulations
and discusses the issue of universality of the critical behavior. 
Sec.~\ref{u23} deals with the analytical results including 
renormalization group approaches.
In Sec.~\ref{u24}, several modifications of the forest-fire model
are discussed.
Some of the most prominent other SOC systems are presented
in Sec.~\ref{u3}. 
Sec.~\ref{u4}  contains concluding remarks.

\section{The self-organized critical forest-fire model}
\label{u2}

\subsection{Critical exponents and scaling relations}
\label{u21}

In this section, we consider some principal properties of the SOC FFM
which will lead us to the definition of critical exponents 
and the derivation of scaling relations between them.
In the following, we restrict ourselves to the case $g=0$.

The mean number $\bar s$ of trees that are destroyed by a
lightning stroke can be calculated as follows. 
During one time step, there are 
$f \rho_t L^d$ lightning strokes in the system and $p \rho_e L^d$
growing trees.
In the steady state, the number of growing trees equals the number of burning
trees, and therefore the mean number of trees destroyed by a lightning
stroke is
\begin{equation}
\bar s = {p \rho_e \over f \rho_t} \simeq {p \over f} {1-\rho_t \over \rho_t}.
\label{eva}
\end{equation}
In the last step, the fire density which is very
small due to time scale separation was neglected.
For small values of $f/p$, the forest density $\rho_t$ approaches a constant
value. If this constant value is less than 1, the second factor on the
right--hand side of Eq.\ (\ref{eva}) is also constant for small $f/p$, and
Eq.\ (\ref{eva}) then represents a power law
\begin{equation}
\bar s \propto (f/p)^{-1}. 
\label{cruz}
\end{equation}
In $d\ge 2$ dimensions, the critical forest density 
\begin{displaymath}
\rho_t^c = \lim_{f/p \to 0} \rho_t,
\end{displaymath}
in fact, must be less than 1, as the following consideration indicates: 
If the critical forest
density were $\rho_t^c=1$ in $d\ge 2$ dimensions, $\rho_t$ would be very 
close
to 1 for small values of $f/p$. Then the largest forest cluster would contain a
non-vanishing percentage of all trees in the system, and the average number
$\bar s$ of trees burned by a lightning stroke would diverge in the limit $L
\to \infty$ with fixed $f/p$, in contradiction to Eq.~(\ref{eva}). In one
dimension, there is no infinitely large forest cluster in the system as long as
$\rho_t <1$, and therefore the critical forest density is $\rho_t^c=1$.
Nevertheless Eq.~(\ref{cruz}) holds 
also in 1 dimension since the forest density approaches its critical value only
logarithmically slowly, as will be shown in Subsection \ref{u23}.
Eq.~(\ref{cruz}) indicates a critical point in the limit $f/p \to 0$. 
Close to this critical point, i.e. if $f \ll p$, there is scaling
over many orders of magnitude.

Let $n(s)$ be the mean number of
forest clusters per unit volume consisting of $s$ trees.
Then the mean forest density is
\begin{equation}
\rho_t =\sum_1^\infty sn(s),
\label{arantxa}
\end{equation}
and the mean number of trees destroyed by a lightning stroke is
\begin{equation}
\bar s = \sum_1^\infty s^2 n(s) /\rho_t. 
\label{marina}
\end{equation}
Since $\lim_{f/p \to 0} \rho_t$ is finite and $\bar s$ diverges as
 $(f/p)^{-1}$, these equations imply that $n(s)$ decreases at least
like
$s^{-2}$ but not faster than $s^{-3}$. As long as the system is not
exactly at the critical point $f/p=0$, i.e. for non-vanishing $f/p$, there must 
be a
cutoff in the cluster size distribution for very large forest clusters.
We conclude that \cite{dro92}
\begin{equation}
n(s)\propto s^{-\tau}{\cal C}(s/s_{\text{max}})
\label{ines}
\end{equation}
with $2\le \tau \le 3$ and
\begin{equation}
s_{\text{max}} \propto (f/p)^{-
\lambda}\propto \bar s^\lambda.
\label{estrella}
\end{equation}
The cutoff function ${\cal C}(x)$ is essentially constant for
$x\le 1$ and
decreases to zero for large $x$. Eqs.~(\ref{marina}) - (\ref{estrella}) 
yield $\bar s
\propto s_{\text{max}}^{3-\tau}$, which leads to the scaling relation
\begin{equation}
\label{nerea}
\lambda=1/(3-\tau).
\end{equation}
In the case $\tau=2$, the right--hand side of Eq.~(\ref{ines}) acquires a
factor $1/\ln(s_{\text{max}})$,
since the forest density given by Eq.~(\ref{arantxa}) 
must not diverge in the limit $f/p\to 0$. The
mean number of forest clusters per unit volume $\sum_1^\infty
n(s)$, therefore, decreases to zero for $f/p\to 0$, and consequently
the forest density approaches the value 1. This situation occurs in one dimension. 

We also introduce the cluster radius $R(s)$ (radius of gyration) which is 
the mean distance of the trees in a cluster from their center of mass. 
It is related to the cluster size $s$ by
\begin{displaymath}
s\propto R(s)^\mu ,
\end{displaymath}
with the fractal dimension $\mu$. 

The correlation length $\xi$ is defined by
\begin{displaymath}
\xi^2 = \frac{2 \sum_1^\infty sn(s) \cdot sR^2(s)}{\sum_1^\infty 
sn(s) \cdot s} \propto (f/p)^{-2\lambda/\mu}, 
\end{displaymath}
We conclude
\begin{equation}
\xi \propto (f/p)^{-\nu} \text{ with }  \nu = \lambda / \mu. 
\label{juana}
\end{equation}

In percolation theory, the {\em hyperscaling relation}
\begin{equation}
\label{carmen}
d=\mu(\tau-1) 
\end{equation}
is satisfied, but it is not satisfied in the SOC
FFM in $d=2$, as first stated in \cite{hen93}, where also
an interpretation of this relation is given: If Eq.\ (\ref{carmen}) is 
satisfied, every box of $l^d \gg 1$ sites 
contains a spanning piece of a large cluster when the
system is at the critical point. In the FFM, there are
at least in $d=2$ many regions which contain no large forest cluster 
(see Fig.~\ref{fig4}), and consequently $d < \mu(\tau-1)$.

The mean forest density $\rho_t$ approaches its critical value $\rho_t^c =
\lim_{f/p \to 0} \rho_t$ via a power law
\begin{equation}
\rho_t^c - \rho_t  \propto (f/p)^{1/\delta}.
\label{angela}
\end{equation}

One can also introduce several dynamical quantities and corresponding
exponents characterizing the 
temporal behavior of the fire (see \cite{cla94}). Here, we mention only the time $T(s_{\text{max}}) \propto (f/p)^{-\nu'}$ which it takes to burn down a cluster of size $s_{\text{max}}$. 

\subsection{Simulation results}
\label{u22}

In this subsection, the simulation results for the critical exponents
of the FFM are presented and discussed.
 
\subsubsection{Critical Exponents}

In one dimension, the critical exponents were determined not only by
computer simulations, but also analytically \cite{dro93}
(See also Subsection \ref{u232}). 
In higher dimensions one has to resort to computer simulations.
The model is simulated most effectively
with a method first proposed in \cite{gra93}, which iterates the 
following rules:
\begin{enumerate}
\item
Choose an arbitrary site in the system. 
If it is not occupied by a tree, proceed  with rule 2.
If it is occupied by a tree, then
ignite the tree and burn down the forest cluster to which the tree belongs. 
While burning the trees, evaluate the properties of the cluster as size,
radius, etc. Proceed with rule 2.
\item
Choose $p/f$ arbitrary sites in the system and grow a tree at all chosen empty
sites. Proceed with rule 1.
\end{enumerate}
By these rules, time scale separation is perfectly realized, 
and Eq.~(\ref{eva}) is satisfied. 
In order to assure that the system is in the steady state, 
a sufficiently large number of time steps have to be discarded 
in the beginning of each simulation. 
In the following, we discuss the simulation results obtained for the 
critical exponents and the critical forest density.

The exponent $\tau$ is determined by measuring the cluster distribution
$n(s)$ or the fire distribution $sn(s)$, which gives $\tau - 1$.
In $d=2$, $\tau \approx 2.15$, as found in \cite{gra93,chr93,hen93,cla94}.
With increasing dimension, $\tau$ increases, too \cite{chr93,cla94}, and 
assumes its mean-field value 2.5 in dimensions larger than 6 \cite{cla94}
(for simulation results see Fig.~\ref{fig5}, and for a complete list of the values see Tab.~\ref{tab1}).

Reliable results for the exponent $\lambda$ can be achieved by considering 
the normalized integrated distribution function 
$P(s) = \int_s^\infty ds' s' n(s')$,
which was first introduced in \cite{gra93}.
By collapsing different curves of $P(s)$ for different values
of $f/p$ , $\lambda$ is found to be $\approx 1.16$ 
\cite{hen93,cla94}. 

The critical forest density is $\rho_t^c \approx 0.41$ in $d = 2$ \cite{gra93,chr93,hen93,cla94}, and it decreases with increasing dimension \cite{chr93,cla94}. Tab.~\ref{tab1} lists the results obtained in \cite{cla94}, where the system size was larger than in \cite{chr93}.
The exponent $1/\delta$ is $\approx 0.5$ in $d = 2$\cite{gra93,chr93,cla94}.

The fractal dimension $\mu$ of the forest clusters is obtained from the
slope of the cluster radius $R(s)$ (see Fig.~\ref{fig6}). In 2 dimensions $\mu
\approx 1.96$ \cite{hen93,cla94}, which is smaller than 2, in contrast to earlier assumptions \cite{dro92,gra93}).
The values in higher dimensions are given in Tab.~\ref{tab1} \cite{cla94}
and seem to approach the mean--field value 4, which is supposedly exact above 6 dimensions.
The hyperscaling relation Eq.~(\ref{carmen}) is definitely violated in
$d=2$ (see \cite{hen93} for an interpretation), 
but cannot be ruled out from simulation results in higher dimensions.

The correlation length $\xi$ is dominated by large clusters and consequently 
large radii. Therefore, the exponent $\nu$ could  be directly 
determined only in $d = 2$ dimensions, with the result $\nu \approx 0.58$ 
\cite{gra93,hen93,cla94}. The temporal analogon to the correlation length is $T(s_{\text{max}}) \propto (f/p)^{-\nu'}$, with $\nu' \simeq 0.61$ in $d = 2$ \cite{cla94}. 

All these simulation results for the critical exponents suggest
that the SOC forest-fire model has
 an upper critical dimension $d_c=6$, 
above which the critical exponents are identical
with those of mean--field--theory, which again is identical to the
mean--field--theory of percolation \cite{chr93}. 
The strongest evidence for this behavior comes from the exponent $\tau$, 
which approaches the percolation value $\tau_{\text{perc}} = 5/2$ for
$d \to 6$ and is indistinguishable from $5/2$ in all simulated
dimensions $d \ge 6$. But also in the other exponents the difference 
between forest--fire and percolation values seems to vanish with increasing
dimension. 

\subsubsection{Universality}
In equilibrium systems, the critical behavior usually depends only on 
properties as dimension and conservation laws, but not on microscopic
details. It can therefore be expected that also the critical behavior of
the SOC FFM is universal under certain changes of the model rules.
This assumption was checked in \cite{cla94}, where the 2D simulations
were carried out for a triangular lattice ($\rho_t^c \approx 34\%$) 
and for a square lattice with 
next--nearest--neighbor interaction ($\rho_t^c \approx 28\%$). 
The critical exponents of these
variations of the model were found to be the same. See, for instance, 
Fig.~\ref{fig7}, where the correlation length for the three models is plotted.
The critical densities are smaller than  
for the square lattice, since the 
fire has more possible paths to spread due to the larger number of neighbors. 

\subsubsection{Immunity}

Another modification of the model rules is obtained by including a non zero
immunity 
$g$ \cite{dro94a}. In contrast to the rule given in Sec.~\ref{u1}, the 
simulations were performed with immune bonds instead of immune sites. 
Not all trees that are neighbors of a 
burning tree catch fire, and consequently the fire does no longer burn forest
clusters but clusters of trees that are 
connected by non immune bonds. A small value of immunity corresponds 
essentially to a change in the lattice symmetry, since the effective 
coordination number is decreased. As we have just seen, a change of the 
lattice symmetry does not affect the critical exponents, which is again 
confirmed by the simulations.
With increasing immunity, the 
forest density increases until the critical forest density becomes $\rho_t^c = 1$ at a critical immunity $g_c$. Here, a new scenario occurs: The forest is completely dense in the limit $f/p \to 0$, and clusters that are destroyed by fire are 
percolation clusters of bond percolation. Consequently, 
the exponents $\tau_c = \tau(g=g_c)$ and 
$\mu_c$ are given by percolation theory, and the threshold  is
$g_c= 1/2$, which is 1 minus the bond percolation threshold.
Since the critical forest density is 1 at $g=g_c$, scaling relation Eq.~(\ref{nerea}) now reads
\begin{equation}
\label{isabel}
\lambda_c=\gamma_c/(3-\tau_c)
\end{equation}
with a new exponent $\gamma_c$.

For finite $f/p$, 
the mean forest density is no longer 1, and one can determine also
the other critical 
exponents $\lambda_c$, $\delta_c$, $\nu_c$, and $\gamma_c$. 
The scaling relations Eq.~(\ref{isabel}) and Eq.~(\ref{juana}) 
are confirmed by the simulations.
Using Eqs.~(\ref{eva}) and (\ref{angela}), one obtains a new scaling relation
$\bar s \propto (f/p)^{-\gamma_c}$ with $\gamma_c = 1 - 1/\delta_c$,
which is also confirmed by the simulations. 

When the immunity is just below its critical 
value $g_c$, a crossover from percolation--like to SOC behavior is observed. 
On length scales smaller than the percolation correlation length
$\xi_{\text{perc}} \propto (g_c - g)^{\nu_{\text{perc}}}$, a system close 
to the percolation threshold  
cannot be distinguished from a system exactly at the percolation threshold. 
As long as $f/p$ is so large that fires do not spread further
than $\xi_{\text{perc}}$, the exponents are identical to those at $g = g_c$. 
When $f/p$ becomes very small, there are fires which spread further
than the percolation correlation length. 
These fires are stopped by empty sites that were created by earlier fires.
This is the same mechanism as 
for small $g$: fires that would spread indefinitely  
if there were no empty sites are stopped 
by empty sites. We conclude that these large fires lead to the critical 
exponents $\lambda$, $\nu$, and $\delta$ that have been observed for $g=0$.
We make the following scaling ansatz for the
correlation length:
\begin{equation}
\xi = (f/p)^{-\nu_c} F\left({g_c-g \over (f/p)^\phi}\right).
\label{dolores}
\end{equation}
It is plausible that the crossover from percolation-like to SOC behavior takes
place when  
$f/p$ becomes so small that the correlation length exceeds the percolation 
correlation length, which suggests that the crossover exponent $\phi$ is 
given by $\phi= \nu_{c}/\nu_{perc}$. 
The scaling function $F(x)$ is constant for small $x$ and is $\propto 
x^{(\nu-\nu_c)/\phi}$ for large $x$. 
Analogous scaling laws hold for $s_{\text{max}}$ and $\rho_t^c-\rho_t$.
Fig.~\ref{fig8} shows the simulation results for the 
scaling function of the correlation length $F(x)$ for  
different values of $g_c-g$. The scaling ansatz Eq.~(\ref{dolores}) is well 
confirmed since all curves coincide. The dashed line represents $F(0)$ as
obtained from the simulations at $g_c$. This crossover is similar to
 crossover phenomena at equilibrium phase transitions, and it should be observed also in higher dimensions. Although the crossover in $\tau$ and $\mu$ vanishes in dimensions larger than 6, there is still a crossover in $\lambda$ and $\nu$, due to the modified scaling relation at $g_c$. In $d=1$, 
the critical immunity is $g_c=0$, and no crossover can take place. 

\subsection{Analytical results}
\label{u23}

In this subsection, we review several analytical approaches to the SOC
forest-fire model. First, we present the mean-field theory (MFT) which leads to
the same exponents as the MFT of percolation. Then, we give several exact
results for the one-dimensional model. Finally, we discuss  attempts to
establish a renormalization group for the forest-fire model.

\subsubsection{Mean-field theory}
\label{u231}

One of the simplest possible analytical approaches to a model is a MFT, which
neglects all correlations in the system and describes it entirely in terms of
densities. The neglect of correlations can be modelled in simulations by
constructing a random-neighbor model, where the nearest-neighbor connections
are chosen randomly at each time step \cite{chr93}. The probability that a
randomly chosen neighbor of a given burning tree is occupied by a green tree is
obviously given by the tree density. In the stationary state, the mean-field
equations are the following \cite{chr93,dro94,dro94c}:
\begin{eqnarray}
\rho_f & = & p\rho_{e}\, , \\
           \rho_f & = & \rho_t\,
\left(f + (1-f)(1 - (1-\rho_f)^{2d})\right) , \\ 
           \rho_{e}+\rho_t+\rho_f  & = & 1\, .
\end{eqnarray}
The first and last equation are exact (see Eq.(\ref{luisa}) 
and Eq.(\ref{conchita})), since
they involve no nearest-neighbor interactions. The second equation includes a
mean-field approximation, since the probability that one or more neighbors of a
given site are burning is given by $(1 - (1-\rho_f)^{2d})$, without taking into
account any correlations.

 From these equations, we find an implicit equation for the fire density alone
\begin{equation}
\rho_f=\left(1-\rho_f\, (1+{1/
p})\right)\left(1-(1-f)\, (1-\rho_f)^{2d}\right).
\end{equation}
In the SOC limit, $p$, $f$, and $\rho_f$ are very small compared to 1, and we
find to leading order
\begin{eqnarray}
\rho_f & = & p(1 - 1/2d) + f/2d + O(p^2, pf, \cdots)\, ,   \\
\rho_e & = & 1 - 1/2d + f/2dp + O(p, f, \cdots) ,  \\
\rho_t & = & 1/2d - f/2dp +  O(p, f, \cdots)\, . 
\end{eqnarray}
The tree density approaches its critical value linearly in $f/p$, leading to
$\delta = 1$. The critical tree density is $\rho_t(f/p=0)=1/2d$, which means
that a burning tree ignites on an average just one other tree. This situation
is identical to MFT of percolation or, equivalently, percolation on a Cayley
tree, where percolation proceeds to each available neighbor with the same
probability and to one neighbor on an average. The cluster size distribution
and the fractal dimension of clusters for this problem are well known
\cite{sta92}. We therefore obtain $\tau = 2.5$ and $\mu = 4$. Scaling
relations Eqs.~(\ref{nerea}) and (\ref{juana}) 
then give $\lambda = 2$ and $\nu = 0.5$. Percolation has
the upper critical dimension 6, and one might expect that the SOC forest-fire
model has the same upper critical dimension, as conjectured in 
\cite{gra93,chr93,cla94} and supported by the simulation
results (see subsection \ref{u22}).

When the limit $p \to 0 $ with $f = 0$ is considered instead of the limit $f
\to 0$ with $p \ll f/p$, the MFT of the forest-fire model also gives a critical
tree density $1/2d$, leading again to the same exponents as MFT of percolation.
In MFT the structural information is lost and thus
MFT cannot see the qualitative difference between the SOC state and the
quasideterministic state with spiral-shaped fire fronts.

\subsubsection{Exact results in one dimension}
\label{u232}

In one dimension, exact results have been obtained in 
\cite{dro93,dro94,pac93}, thus
proving analytically that non conservative models can indeed show SOC. Here, we
give an intuitive derivation of the results: Consider a string of $k \ll
p/f$ sites. This string is too short for two trees to grow during
the same time step. Lightning does not strike this string before all of its
trees are
grown. Since we are always interested in the limit $f/p \to 0$, the following
considerations remain valid even for strings of a very large  size.
Starting with a completely empty string, it passes
through a cycle which is illustrated in Fig.~\ref{fig9}.
During one time step,
a tree grows with
probability $p$ on any site. After some time, the string is
completely occupied by trees. Then the forest in the neighborhood of the string
will also be quite dense. The trees on the string are part of a forest cluster
which is much larger than $k$. Eventually that cluster becomes so large that it
is struck by lightning with a non-vanishing probability. Then the forest cluster
burns down, and the string again becomes completely empty.

This consideration allows us to write down rate equations for the states of the
string. In the steady state, each configuration of trees is generated as often
as it is destroyed. Let $P_k(m)$ be the probability that the string is occupied
by $m$ trees. Each
configuration which contains the same number of trees has the same probability.
A configuration of $m$ trees is destroyed when a tree grows at one of the empty
sites, and is generated when a tree grows in a state consisting of $m-1$
trees.
The completely empty state is generated when a dense forest burns down. Since
all trees on our string burn down simultaneously, this happens each time when a
given site of the string is set on fire. 
This in turn happens as often as a new tree
grows at this given site, i.e. with probability $p\, (1-\rho_t)$ per time step.
We therefore have the following equations,
\begin{eqnarray}
p k P_k(0) & = & p\, (1-\rho_t), \\
p(k-m)P_k(m) & = & p\, (k-m+1)P_k(m-1)\;
\text{ for}\; m \neq 0,k\, .
\end{eqnarray}
We conclude
\begin{eqnarray}
P_k(m) & = & (1-\rho_t)/(k-m)\; \text{ for}\;m<k,  \\
P_k(k) & = & 1-(1-\rho_t)\sum_{m=0}^{k-1}1/(k-m)  \nonumber \\ 
& = & 1-(1-\rho_t) \sum_{m=1}^k 1/m.
\end{eqnarray}
These last two equations contain a wealth of information: cluster size
distribution, hole distribution, growth velocity etc.

A forest cluster of size $s$ is a configuration of $s$ neighboring trees with
an empty site at each end. The size distribution of forest clusters
is consequently given by 
\begin{equation}
n(s) ={P_{s+2}(s)\over
{s+2 \choose s}} =  {1-\rho_t\over (s+1)(s+2)} \simeq
(1-\rho_t)s^{-2}.
\label{asuncion}
\end{equation}
This is a power law with the critical exponent $\tau=2$.
The size distribution of fires is
$\propto sn(s)\propto s^{-1}$.
Fig.~\ref{fig10} shows the numerical result for the fire distribution $sn(s)$.
It agrees perfectly with Eq.~(\ref{asuncion}) in the region $s <
s_{\text{max}}$.

$s_{\text{max}}$, introduced in Eqs.~(\ref{ines}) and (\ref{estrella}) 
is the characteristic size (in one dimension length)
where the power law $n(s)\propto s^{-2}$ breaks down.
We calculate
$s_{\text{max}}$ from the
condition that a string of size $k\le s_{\text{max}}$ is not
struck by lightning until all trees are
grown. When a string of size $k$ is completely empty at time
$t=0$, it will be occupied by $k$ trees after
\begin{displaymath}
T(k)=(1/p)\sum_{m=1}^k 1/m \simeq \ln(k)/p
\end{displaymath}
 time-steps on an
average. The mean number of trees after $t$ time-steps is
\begin{displaymath}
m(t)=k[1-\exp(-pt)].
\end{displaymath}
 The probability that lightning
strikes a string of size $k$ before all trees are grown is
\begin{displaymath}
f\sum_{t=1}^{T(k)} m(t)\simeq (f/p)k(\ln(k)-1) \simeq
(f/p)k\ln(k).
\end{displaymath}
We conclude
\begin{displaymath}
s_{\text{max}}\ln(s_{\text{max}})
\propto p/f \; \text{ for large $p/f$}\, , \nonumber
\end{displaymath}
leading to $\lambda=1$.

Next we determine the relation between the mean forest density
$\rho_t$ and the parameter $f/p$.
The mean forest density is given by
\begin{eqnarray*}
\rho_t & \simeq & \sum_{s=1}^{s_{\text{max}}} sn(s) \\
     & = & (1-\rho_t)\sum_{s=1}^{s_{\text{max}}}{s\over
           (s+1)(s+2)}\\
     & \simeq & (1-\rho_t) \ln(s_{\text{max}}).
\end{eqnarray*}
Thus
\begin{displaymath}
{\rho_t \over 1-\rho_t}  
\simeq  \ln(s_{\text{max}}) \simeq \ln(p/f)\, \; \text{
for large $p/f$}. \nonumber
\end{displaymath}
The forest density approaches the value 1 at the
critical point. This is not
surprising since no infinitely large
cluster exists in a one-dimensional
system as long as the forest is not completely dense.

The exponents characterizing the size distribution of forest clusters and fires
remain the same when the fire is allowed to jump over holes of several empty
sites \cite{dro94b}, proving the 
universality of the critical exponents. A change
in the range of the interaction corresponds to a change in lattice symmetry in
higher dimensions, where universality also has been found (see subsection
\ref{u22}).

The values of the exponents change, however, when the rule for tree growth is
modified. When tree growth becomes more deterministic, the critical behavior
eventually breaks down, and the system becomes synchronized \cite{dro95}. To
illustrate this, let us look at the case of completely deterministic tree
growth, where an empty site turns to a tree exactly $T$ time steps after the
site has become empty. Neighboring sites both of which happen to be occupied 
by a
tree burn down during the same fire. They consequently turn to trees
simultaneously and burn down together simultaneously for all future times, i.e.
they are synchronized. After some time, the system will consist of
synchronized blocks which are so large that they are struck by lightning before
the neighboring blocks become occupied by trees.

As an intermediate case, let us next consider a tree growth rule where the life
time distribution of empty sites is a constant $1/T$ over a time $T$. An initially empty string of
$k$ sites then grows the same number of trees during each time step,
irrespective of the number of occupied sites. The string consequently becomes
occupied after a finite time. The above--mentioned 
deterministic model is just a coarse-grained 
version of this model, and we therefore expect that the system is
dominated by large fires. Due to stochastic tree growth, however, there exist
also small forest clusters. At small scales $k$, we find that  $P_k(m)$ has the
same value for all $m < k$, leading to $n(s) \propto s^{-3}$. The simulation
results Fig.~\ref{fig11} show this power law for small cluster sizes, and the
expected peak at large clusters, indicating synchronization. When the
distribution of life times of empty sites has a power-law tail, the system
shows SOC behavior with an exponent which depends on the exponent
characterizing the tail \cite{dro95}.

\subsubsection{Renormalization group approach}
\label{u233}

In equilibrium critical phenomena, the invariance of the system under a change
of length and time scale together with a partial trace operation leads 
to a renormalization group (RG) approach. A RG
not only yields values (possibly approximate, if one has to use perturbation 
theory) for the critical exponents in low
dimensions, but also information on the upper critical dimension and on
universality classes.

The separation of time scales and the avalanche structure of the forest-fire
model and other SOC models make it difficult to find an appropriate RG
formalism for these systems. 
Two approaches which have been taken so far
\cite{lor95,pat94} do not yet discriminate between the two limits
$\lim_{p \to 0} \lim_{f \to 0}$ and $\lim_{f \to 0}$ with $p \ll (f/p)^{\nu'}$.
The real-space RG in \cite{lor95}, which neglects correlations,
yields the correct exponents in one dimension, and good approximations in two
dimensions. 
In \cite{pat94}, the FFM is mapped on a field theory, but 
the subsequent calculations contain severe approximations. 
In a third approach \cite{bot95}, the forest-fire model is also mapped on a
field theory. However, the renormalization 
in the limit of double time scale separation remains a problem. 

\section{Modifications of the forest-fire model}
\label{u24}

The forest-fire model can be modified in many ways, and some of these
modifications show very interesting behavior. A version with conserved tree
density is suggested in \cite{dro94} and studied in \cite{cla95}. After each
fire, all trees which have been burnt are regrown at randomly chosen 
empty sites. For
tree densities smaller than $\rho_t^c$, this model has only finite fires and
can be mapped on the SOC forest-fire model by finding the corresponding value
of $f/p$. For densities between $\rho_t^c \simeq 0.41$ and a second critical
density $\rho_t^{c2} \simeq 0.435$, the system shows critical behavior with a
power-law size distribution of fires and of tree clusters. Both the cutoff in
cluster size and the correlation length diverge with some power of the system
size (see Fig.~\ref{fig12}). The exponents depend on the density. Thus
one has the interesting situation that criticality is not confined to a point
but exists in a finite intervall.
For even
higher average density, the system splits into several subphases with 
different densities
(Fig.~\ref{fig13}). The subphase with highest density contains an infinite
cluster. The number of subphases which can be sustained 
depends on the system size, their shape 
on the boundary conditions. Another modification of this model, where
each tree is grown immediately after it has been burnt, shows both critical
behavior and spiral-shaped fire fronts \cite{cla96}.

A deterministic version of the forest-fire model with continuous tree growth
has been suggested in \cite{che90}. The variable ``tree height'' is increased
globally and very slowly. Trees higher than a certain threshold catch fire and
ignite all neighbors above a second, smaller, threshold. The height of the tree
after the fire is a function of the height before the fire. \cite{soc93}
studies a generalized version and shows that this model is not critical, but
shows periodic behavior or finite fires, depending on the value of a
parameter. 
Another continuous model, which additionally includes energy
diffusion, shows small or large fires, depending on the parameters, and the two regimes are separated by a critical point. 
\cite{cha95}.

\section{A selection of other SOC models}
\label{u3}

Many systems have been introduced as examples for SOC. For part of these
systems, the evidence is based on numerical data for relatively small system
sizes with little analytical understanding of the 
origin of the critical behavior. In this section, we
discuss three of the most thoroughly studied SOC models and compare them with
the
forest-fire model. Although the term SOC has been used in many different
contexts, we restrict ourselves to systems with slow driving (energy input) and
dissipation events which are instantaneous on the time scale of driving. The
size distribution of dissipation events obeys a power law.
Further examples for SOC include the dynamics of magnetic domain patterns 
\cite{che90a,bak92}, cloud formation \cite{nag92}, evolution of 
populations \cite{bak89,sne93}, erosion \cite{tak92}, 
fracture \cite{ber95}, fragmentation \cite{odd93},
the Bean critical state in type II superconductors \cite{tan93},
depinning transitions of interfaces, charge density waves or 
superconducting flux lattices \cite{pac95}, and many more. 
Although not treated in this section, these examples underline the 
importance of SOC as a crossdisciplinary
subject with applications in physics, biology, chemistry and geology.

\subsection{Sandpile model}
\label{u31}

The sandpile model is the prototype for SOC \cite{bak87,bak88}. Sand grains are
dropped at random on the sites of a lattice. When the
number of grains on a lattice
site (``height model'') or the difference in the grain number between 
neighbors
(``slope model'') exceeds a certain threshold, the site topples and
redistributes its grains amongst 
the nearest neighbors. When a neighbor thereby is lifted
above the threshold, it topples too, and the avalanche continues until all
sites are below or at the threshold. Then a new grain of sand is added at
random. In the stationary state, the mean number of added grains must equal the
mean number of grains leaving the system at the edge, and  there
must exist avalanches spreading over a distance of the order of the system
size. The sandpile therefore is in a critical state where sand grains are
redistributed on all scales. For the height model, several exact results have
been obtained \cite{dha90a}. The critical state is robust with respect to a
variety of
changes, but is usually destroyed when the local conservation law is violated.
Real sandpiles behave differently, mainly because large avalanches have large
inertia and are not immediately stopped when the slope becomes small.

In contrast to the forest-fire model, the sandpile model has only two separated
time scales. The condition that avalanches relax before new grains of sand are
added, corresponds to the condition that a fire is extinguished before new
trees grow. In the sandpile model, large avalanches occur due to the local
conservation of sand grains, in the forest-fire model large fires occur in the
limit $f/p \to 0$. It has been pointed out in \cite{lor95} that the sandpile
model can be simulated exactly at the critical point, while the forest-fire
model is only close to its critical point and has therefore a relevant
parameter. However, in any realistic physical situation, the driving rate for
the sandpile model can never be exactly equal to zero, resulting also in a
relevant parameter and an upper cutoff in avalanche size. Above that size,
avalanches overlap.

\subsection{Earthquake model}
\label{u32}

The earthquake model \cite{ola92} is a continuous, non conservative variant of
the sandpile model. It can be derived from a 2D spring-block model for the
motion of tectonic plates (Burridge-Knopoff model). The force on all sites (``blocks'') is increased
uniformly and very slowly. When the force at a given site  exceeds a threshold,
the block moves to its equilibrium position, and the force on the corresponding
site is reset to zero. The force on the 4 nearest neighbors is increased by a
fixed percentage of the released force. This percentage depends on the ratio of
the spring constants in the model, and is usually smaller than 1/4, i.e. the
model has no conservation law. The system is not driven as long as there are
forces above the threshold and an ``earthquake'' is going on. The size
distribution of earthquakes is found to obey a power-law, in agreement with the famous Gutenberg-Richter law for the size distribution of real earthquakes
. The exponent depends
continuously on the degree of conservation, and the critical behavior seems to
persist even in the limit where almost no force is transmitted to the
neighbors. The size of the largest avalanche diverges with diverging system
size, but so slowly that it covers only a vanishing part of the system
\cite{jan93,gra94}. When the global driving is replaced by local driving, 
when inhomogeneities such as defects are present, or when the 
open boundary 
conditions are replaced by boundary conditions,
the critical state is destroyed \cite{mid95}. The authors of \cite{mid95} show
that the critical behavior is related to a partial synchronization of
neighboring sites. However, the origin of the critical behavior of the
earthquake model, and in particular the dependence of the critical exponents on
the degree of conservation, is relatively poorly understood. A similar behavior
has been found so far only in one other model, namely the above-mentioned
version of the forest-fire model with tree conservation \cite{cla95}.

In this context, another earthquake model that is also derived form the Burridge-Kopoff model, is worth mentioning \cite{car89}. Like the above model, it is  deterministic and continuous, but it includes additionally effects of inertia. It shows a power-law size distribution over a certain range of earthquake sizes, but the very large earthquakes occur more frequently than expected from a power-law distribution, and in periodic time intervals. From the evaluation of earthquake data, it is not obvious, which of the two models is more appropriate. 

\subsection{Evolution model}
\label{u33}

A particularly simple SOC model is the so-called ``evolution model''
\cite{sne93}. Each site in a one-dimensional chain is assigned a random number
between 0 and 1 (``fitness''). The probability for a ``mutation'' depends
exponentially on the fitness, and in the zero-temperature limit always the site
with smallest fitness is mutated. This site is assigned a new random number
between 0 and 1. Since a mutation at one site affects the fitness of the
neighbors, the two nearest neighbors are also assigned new random numbers.
After some time, only a vanishing percentage of all sites have a fitness below
a threshold $\simeq 0.67$. A mutation of a site at the threshold releases an
avalanche of mutations which is stopped when no site is below the threshold any
more. The size distribution of these avalanches obeys a power law. A review of
the properties of this and related  models and of some exact results is given
in \cite{pac95}.

Reference \cite{pac95} mentions also a less spectacular application of this model, 
namely the depinning of interfaces, which usually occurs when a driving force 
exceeds a certain threshold. When not the driving force, but the interface 
velocity is chosen to be the parameter,  the depinning transition becomes SOC 
in the limit of zero velocity \cite{hav91,sne92,zai92}, which is again a 
separation of time scales. 

\section{Conclusions}
\label{u4}

The main part of this review was devoted to the forest-fire model
as a simple example for driven, dissipative systems with many degrees of 
freedom. Depending on the value of
the parameters, the model shows fundamentally different structures. 
One finds a quasi-deterministic state with spiral waves, percolation-like
behavior and, in
particular, a self--organized critical state in the limit of double 
time scale separation. The properties of the SOC state have been analyzed
numerically and analytically. The critical behavior has been shown
to be very robust with respect to a variety of changes of the rules
of the model.

The notion of SOC has originally been introduced as possible
explanation for the ubiquity of fractal structures and $1/f$-noise in 
nature. Although this is
a fascinating hypothesis, SOC seems to account only for the existence
of part of these phenomena. Other
extensively studied mechanisms producing fractals are diffusion-limited 
aggregation
(DLA), kinetic roughening, and chaos and turbulence.

All SOC models presented in this paper have slow driving and avalanche--like 
dynamics. However, not all systems that show these two features are SOC. Besides a power--law size distribution of avalanches,  such 
systems might also have many small avalanches which release only little energy,
or only large avalanches which release a finite part of the system's energy, 
or some combination of both. SOC systems are naturally at the 
critical point, due to a conservation law (sandpile model), a second time 
scale separation (forest-fire model),  a competition between open boundary 
conditions and the tendency of 
neighboring sites to synchronize (earthquake model), or due to the slow 
driving alone (evolution model). However, the critical behavior often 
breaks down when details of the model rules are changed, and is replaced by 
some other scenario. As examples we mentioned the forest--fire model with 
deterministic tree growth and the earthquake model with modified boundary 
conditions. 

The forest--fire model is closely related to excitable media. So far,
spiral waves, target patterns and chaotic behavior have been found
and investigated in excitable media. By investigating the appropriate
parameter region one should also be able to find percolation-like 
behavior and the
self--organized critical state, which exist in the FFM.
The SOC state should occur
whenever there is spontaneous excitation that spreads very fast compared 
to the recovery time (e.g. fatal diseases, that occur seldom, but spread
rapidly),  provided that the probability distribution for the recovery time
is not too narrow. 
 
After many years of studying formation of structure in non-equilibrium systems,
one has got a glimpse at the mechanisms producing the overwhelming
variety and complexity of structures surrounding us in nature,
and there are certainly still many exciting phenomena to be discovered.

\newpage
\begin{figure}
\vskip 1cm
\psfig{file=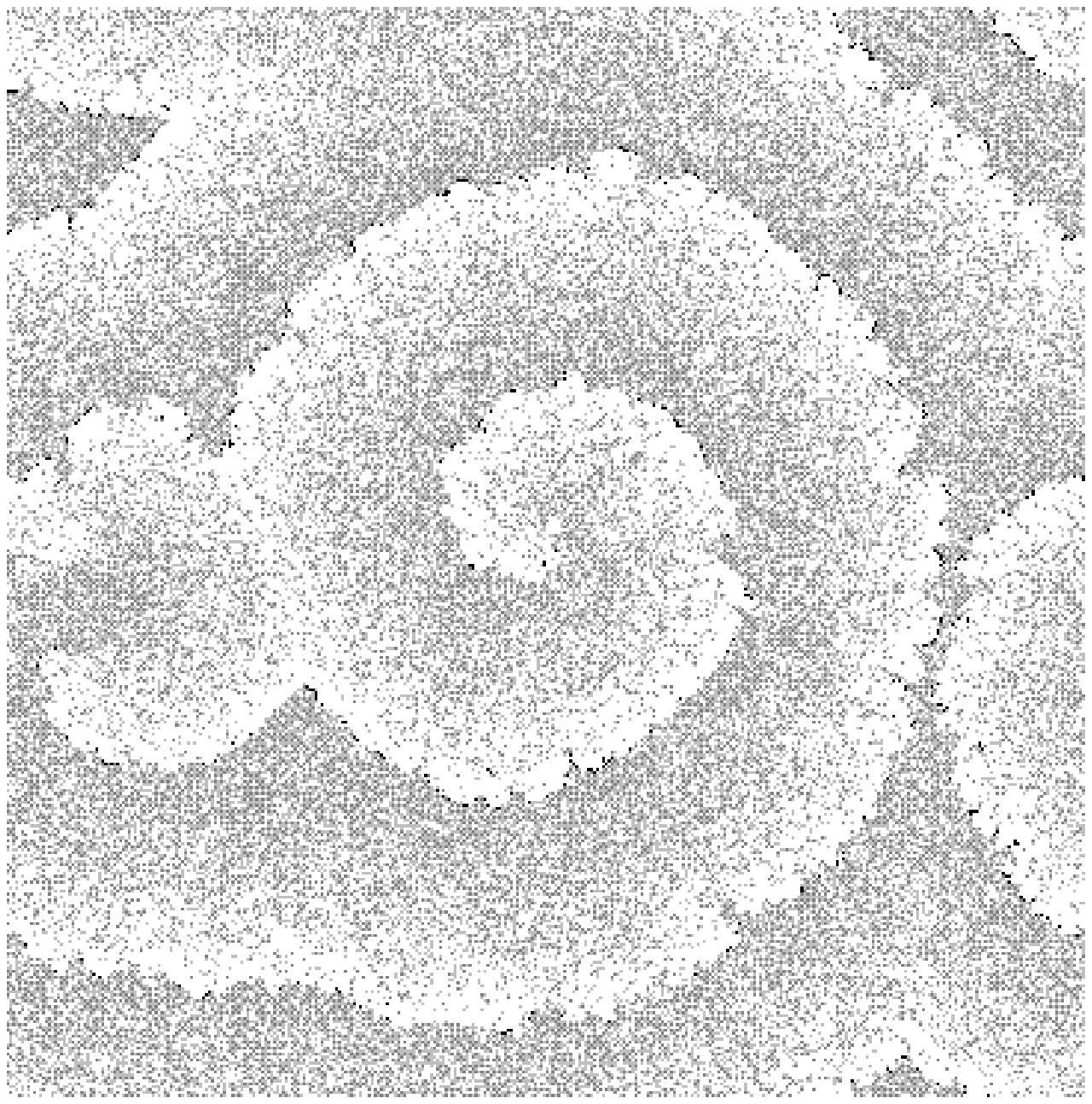,height=6in,angle=0}
\vskip 1cm
\caption{Snapshot of the Bak et al. forest-fire model in the steady state
for $p=0.005$ and $L=800$. Trees are grey, burning trees are black, and empty sites are white. }
\label{fig1}
\end{figure}

\begin{figure}
\vskip -1cm
\psfig{file=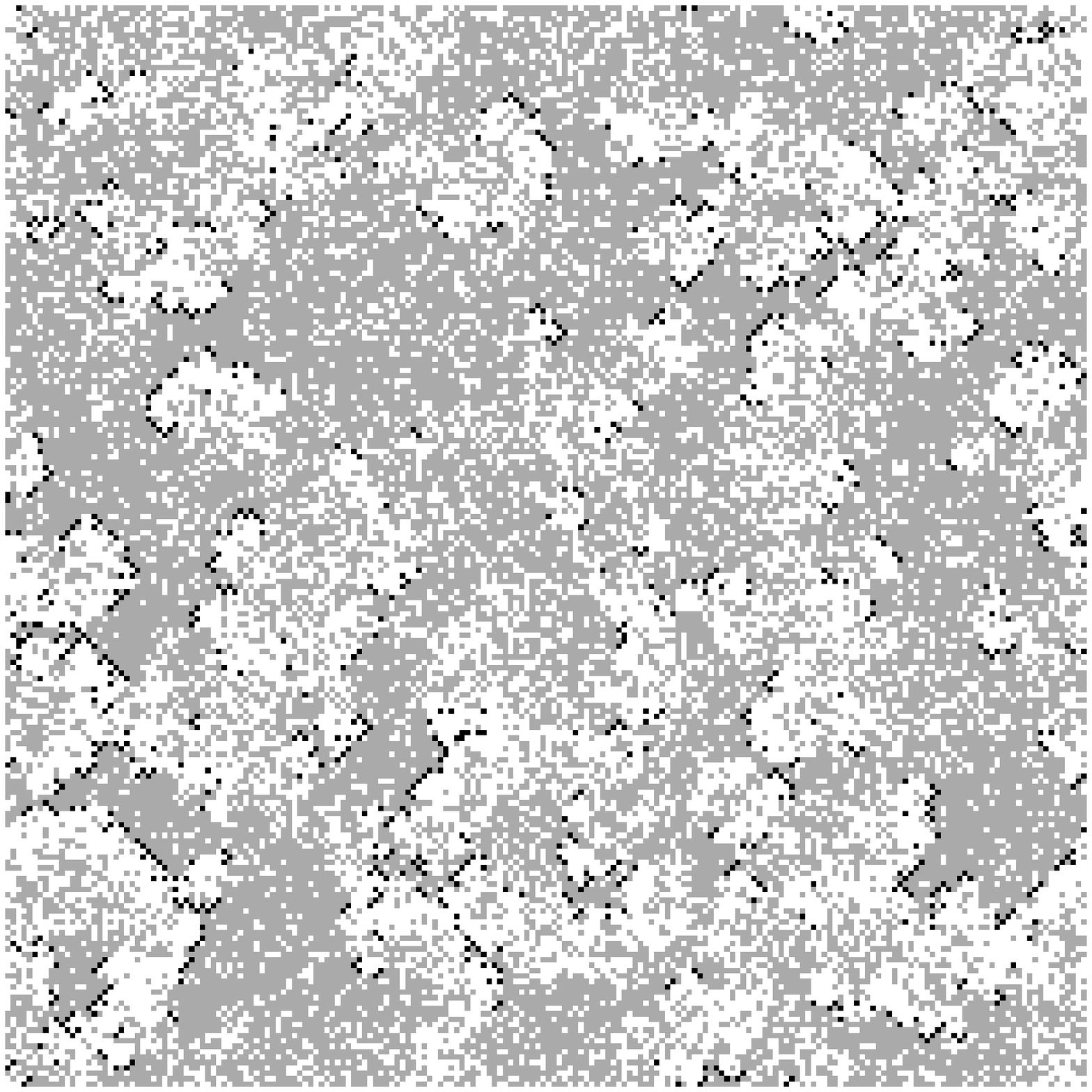,height=4in,angle=0}
\vskip 1cm
\caption{Snapshot of the forest-fire model far below the critical immunity
for $g=0.2$, $p=0.05$, and $L=200$. Trees are grey, empty sites are
white, and burning trees are black.}
\label{fig2}
\end{figure}

\begin{figure}
\vskip -3cm
\psfig{file=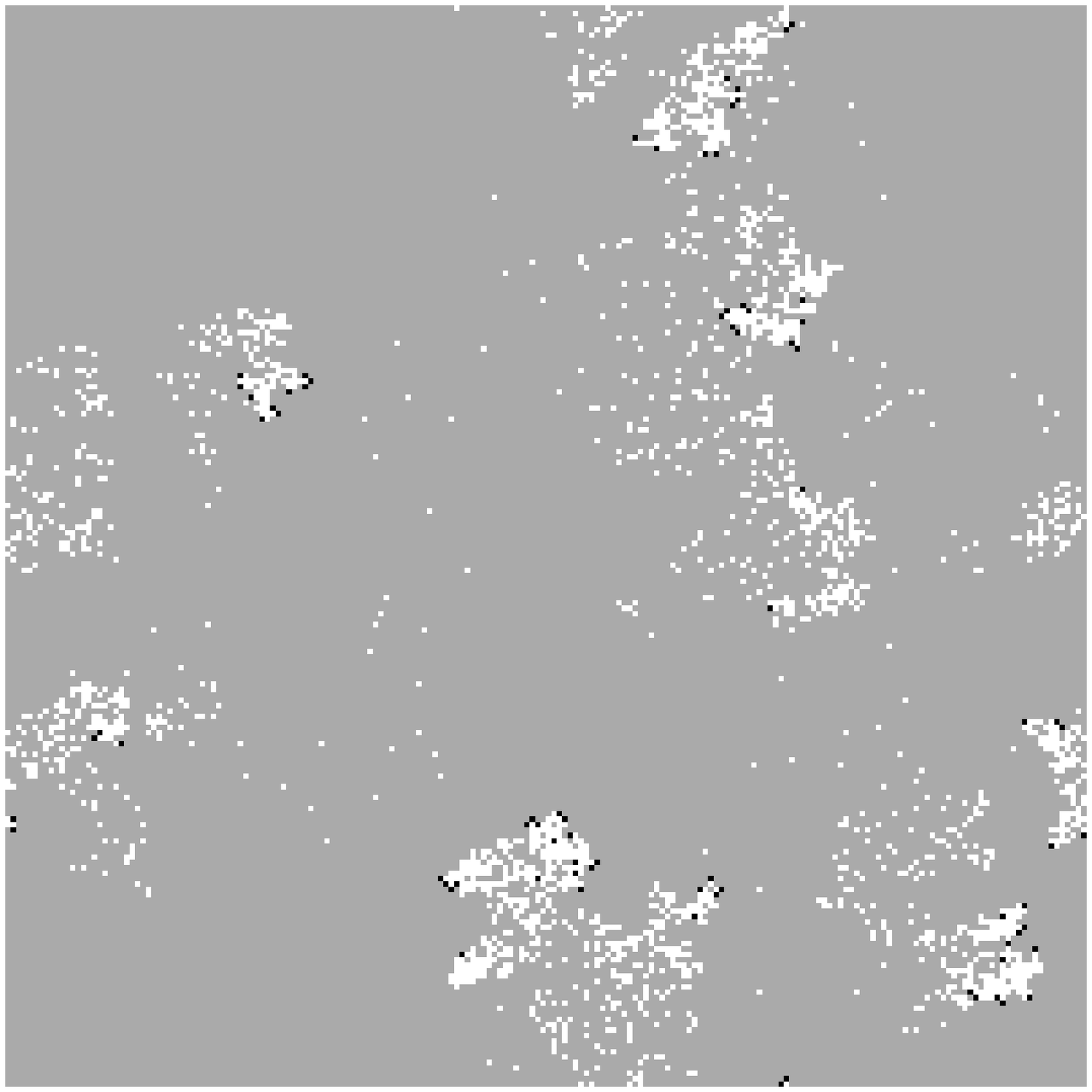,height=4in,angle=0}
\vskip 1cm
\caption{Snapshot of the forest-fire model near the critical immunity
for $g=0.48$, $p=0.05$, and $L=200$. Trees are grey, empty sites are
white, and burning trees are black.}
\label{fig3}
\end{figure}

\begin{figure}
\psfig{file=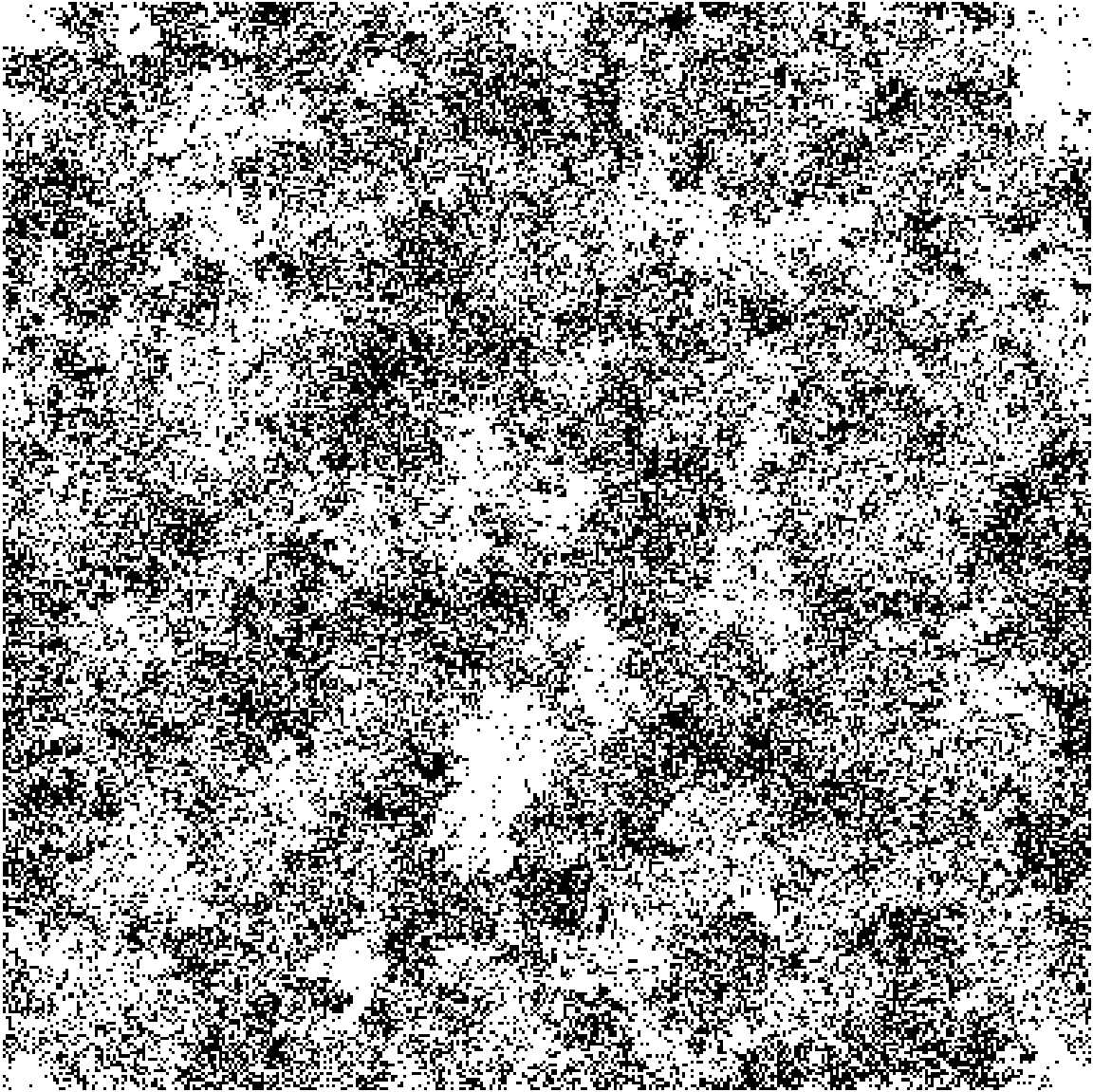,height=6in,angle=0}
\vskip 1cm
\caption{Snapshot of the SOC state in 2 dimensions. Trees
         are black, empty sites are white. The parameters are $L = 1024$ 
         and $f/p = 1/500$.}
\label{fig4}
\end{figure}

\begin{figure}
\psfig{file=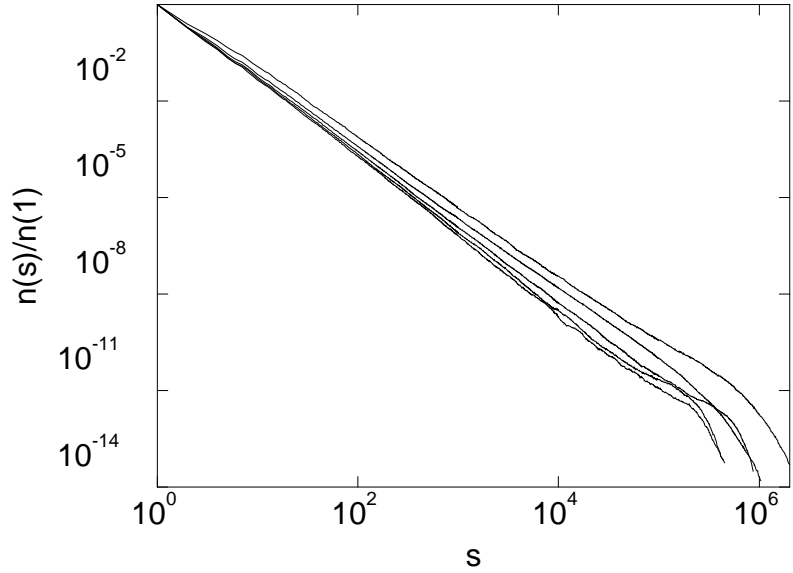,height=3in,angle=-90}
\vskip 1cm
\caption{Normalized cluster size distribution $n(s)/n(1)$ for $d = 2$ 
         to 6 dimensions. The values of $f/p$ are 1/32000, 1/2000, 
         1/1000, 1/250 and 1/125 from right to left. The exponent $\tau$ 
         is given by the negative slope.}
\label{fig5}
\end{figure}

\begin{figure}
\psfig{file=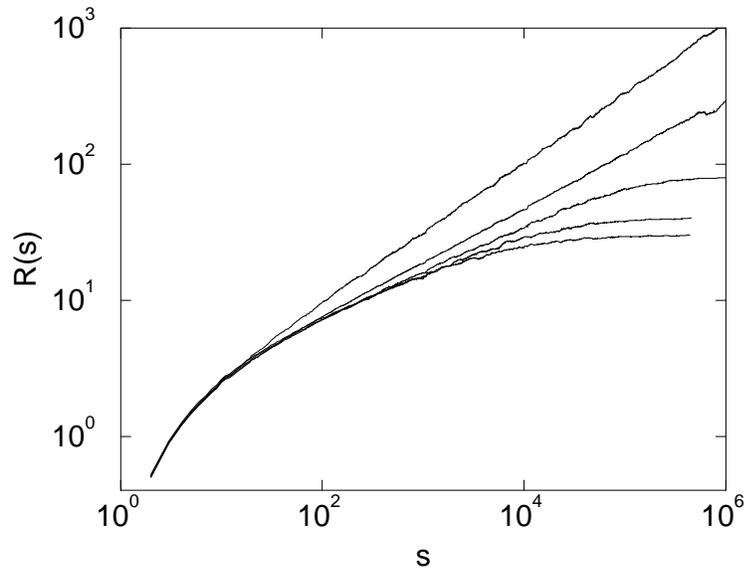,height=3in,angle=-90}
\vskip 1cm
\caption{Cluster radius $R(s)$ in 2 to 6 dimensions (from left to right). 
         The inverse slope yields the fractal dimension $\mu$.}
\label{fig6}
\end{figure}

\begin{figure}
\psfig{file=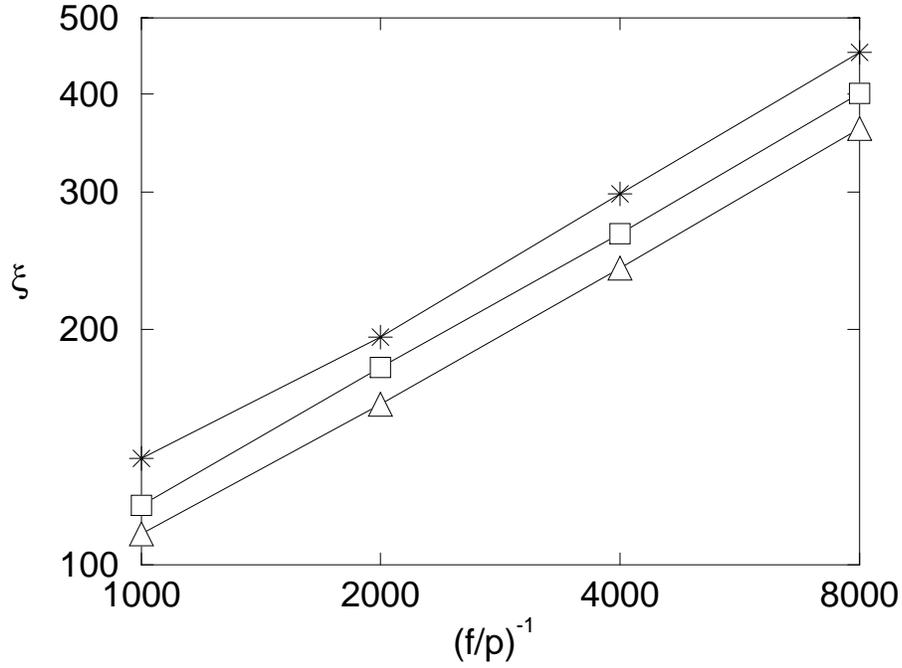,height=3.5in,angle=-90}
\vskip 1cm
\caption{The correlation length $\xi$ as function of $(f/p)^{-1}$. The
         slope yields the critical exponent $\nu$. ($\Box$ = square lattice,
         $\triangle$ = triangular lattice, $\ast$ = next--nearest--neighbor 
         interaction.)}
\label{fig7}
\end{figure}

\begin{figure}
\psfig{file=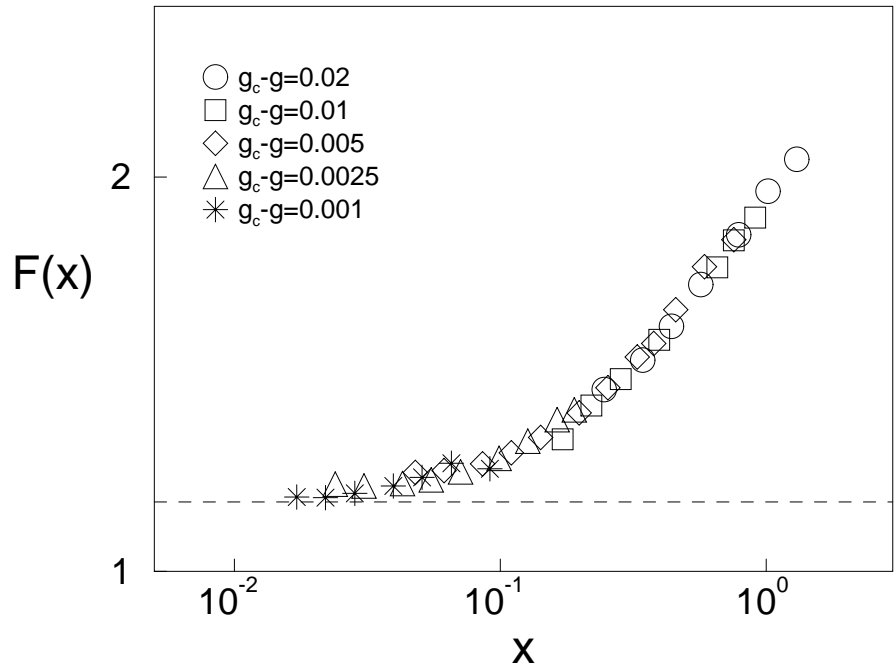,height=3.5in,angle=-90}
\vskip 1cm
\caption{Crossover scaling function $F(x)$ for the correlation
length for different values of the immunity. The dashed line
represents $F(0)$ as obtained at $g=g_c$.}
\label{fig8}
\end{figure}

\phantom{a}
\begin{figure}
\vskip 15cm
\caption{Dynamics on a string of $k = 4$ sites. Trees are black, empty sites
are white}
\vskip -20.5cm
\psfig{file=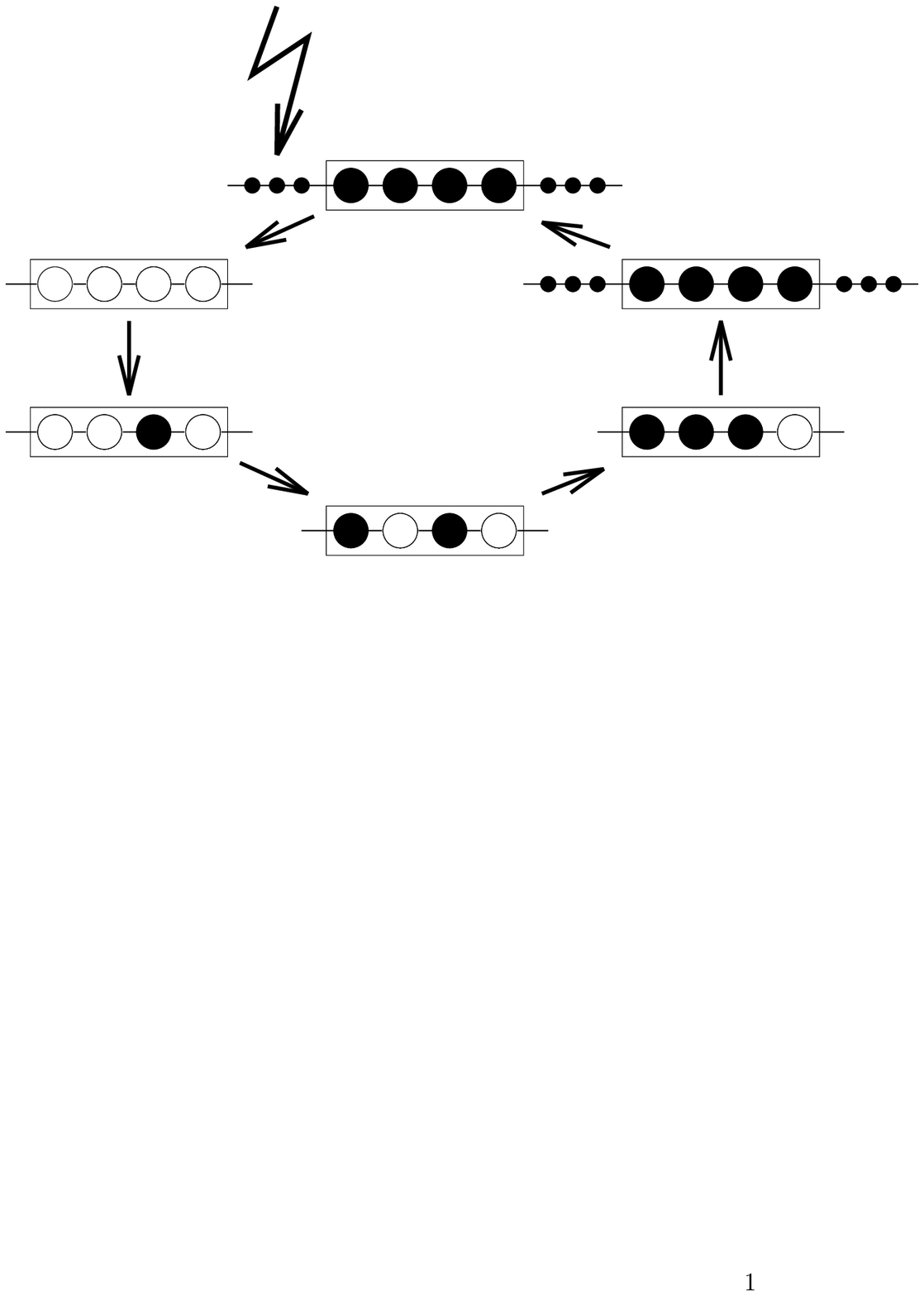,height=9in,angle=0}
\label{fig9}
\end{figure}

\begin{figure}
\psfig{file=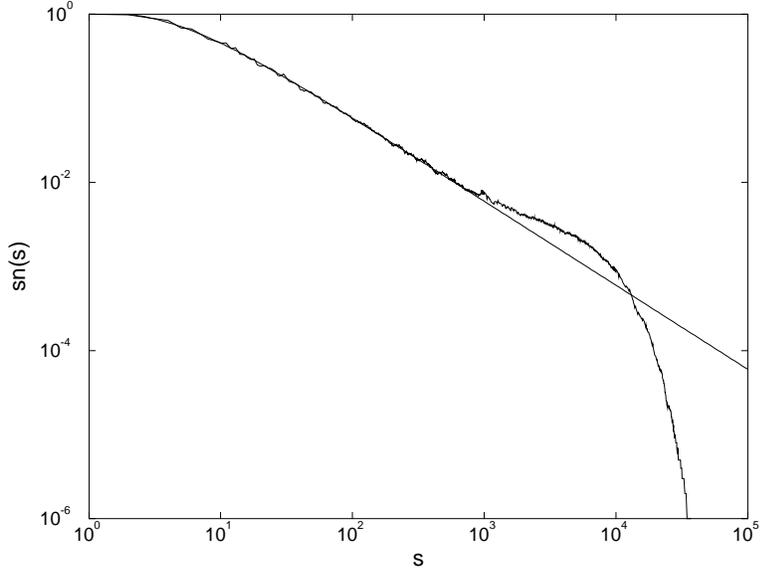,height=3in,angle=-90}
\vskip 1cm
\caption{Size distribution of the fires for $d = 1$, $f/p=1/25000$ and $L=2^{20}$.
The smooth line is the theoretical result which is valid for cluster 
sizes $\le s_{\text{max}}$.}
\label{fig10}
\end{figure}

\begin{figure}
\vskip 1cm
\psfig{file=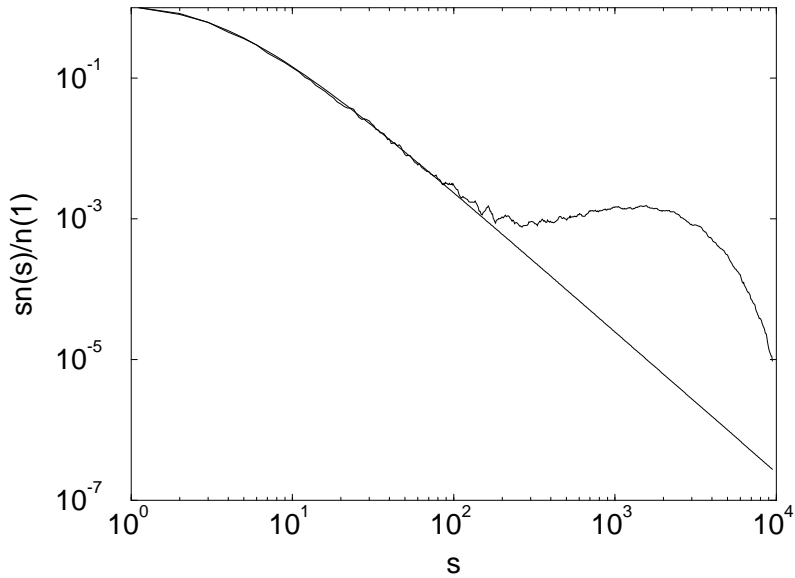,height=3in,angle=-90}
\vskip 1cm
\caption{Normalized size distribution of fires in a version where the life time distribution of empty sites is a constant function $P(\tau)=1/100$ for times $\tau < 100$. The parameters are $L=10000$ and $f=0.00001$. The smooth line is the theoretical
result, which is valid for $s \le s_{\text{max}}$.}
\label{fig11}
\end{figure}

\begin{figure}
\psfig{file=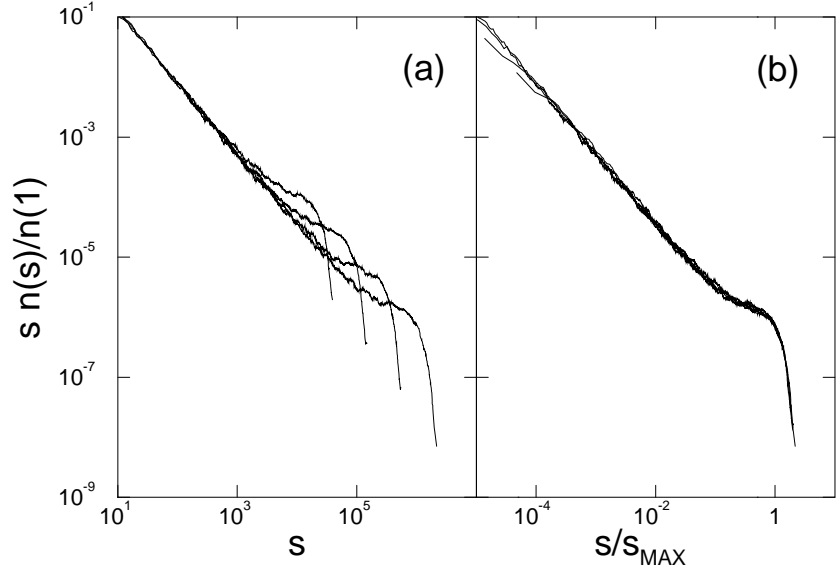,height=3in,angle=-90}
\vskip 1cm
\caption{Normalized size distribution of fires in the model with tree 
conservation, for 
$\rho = 0.43$ and $L$ = 512, 1024, 2048, 4096, (a) before and 
(b) after rescaling.}
\label{fig12}
\end{figure}

\begin{figure}
\psfig{file=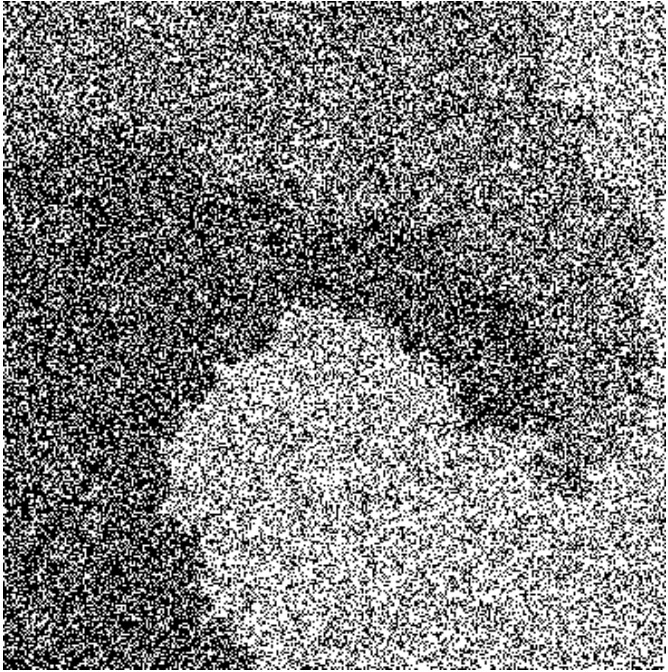,height=3.5in,angle=0}
\vskip 1cm
\caption{Stationary state in the model with tree 
conservation for $\rho = 0.50$, $L = 2048$ and 
         absorbing boundary conditions
         (trees are black, empty sites are white).}
\label{fig13}
\end{figure}

\clearpage
\widetext
\begin{table}
\squeezetable
\caption{Numerical results for the critical exponents in 1 to 8
         dimensions (${}^* = $ with logarithmic corrections, ${}^\dagger = $ 
         calculated from scaling relations), taken from [26]. 
         The exponents with index
         ``perc'' are those of percolation theory [33].} 
\begin{tabular}{ldddddddd}
$d$                             
& 1  & 2  & 3 & 4 & 5  & 6  & 7 & 8 \\
\tableline

$L$                             
& $2^{20}$ & 16384 & 448 & 80 & 32 & 20 & 12 & 8 \\
\tableline

$\tau$                          
& 2       & 2.14(3) & 2.23(3) & 2.36(3) 
& 2.45(3) & 2.50(3) & 2.50(3) & 2.50(3) \\

$\tau_{\text{perc}}$
& 2    & 2.05 & 2.18 & 2.31
& 2.41 & 2.5  & 2.5  & 2.5  \\ 

$\lambda$ 
& $1^*$              & 1.15(3)            & 1.30(6)        
& 1.$56(8)^\dagger$  & 1.$82(10)^\dagger$ & 2.$01(12)^\dagger$ 
& 2.$01(12)^\dagger$ & 2.$01(12)^\dagger$ \\

$1/\delta$                      
& $0^*$ & 0.48(2) & 0.55(12) & -
& -     & -       & -        & - \\

$\rho_t^c$                      
& 1        & 0.4081(7) & 0.2190(6) & 0.146(1) 
& 0.111(1) & 0.090(1)  & 0.076(1)  & 0.066(1) \\  

$\mu$          
& 1      & 1.96(1) & 2.51(3) & 3.0  
& 3.2(2) & -       & -       & -   \\

$\mu_{\text{perc}}$ 
& 1    & 1.90 & 2.53 & 3.06 
& 3.54 & 4    & 4    & 4    \\ 

$\nu$
& $1^*$             & 0.58              & 0.$52(3)^\dagger$ 
& 0.$53(3)^\dagger$ & 0.$57(7)^\dagger$ & -                 
& -                 & - \\
$\nu'$  
& $1^*$              & 0.58               & 0.$64(6)^\dagger$  
& 0.$78(8)^\dagger$  & 0.$92(10)^\dagger$ & 1.$04(11)^\dagger$ 
& 1.$05(11)^\dagger$ & 1.$05(11)^\dagger$ \\
\end{tabular}
\label{tab1}
\end{table}

\begin{references}
\bibitem{man83}  B. B. Mandelbrot,
                 in {\em The Fractal Geometry of Nature}, 
                 (Freeman, New York, 1983).
\bibitem{bun91}  A. Bunde and S. Havlin, 
                 in {\em Fractals and Disordered Systems},
                 ed. by A. Bunde and S. Havlin, 
                 (Springer, New York, 1991).
\bibitem{pre78}  W. H. Press,
                 Comm. Astrophys. {\bf 7}, 103 (1978).
\bibitem{dut81}  P. Dutta and P. M. Horn,
                 Rev. Mod. Phys. {\bf 53}, 497 (1981).
\bibitem{bak87}  P. Bak, C. Tang and K. Wiesenfeld, 
                 Phys. Rev. Lett. {\bf 59}, 381 (1987).
\bibitem{bak88}  P. Bak, C. Tang and K. Wiesenfeld, 
                 Phys. Rev. A {\bf 38}, 364 (1988).
\bibitem{bak94}  P. Bak and M. Creutz, 
                 in {\em Fractals in Science},
                 ed. by A. Bunde and S. Havlin, 
                 (Springer, Berlin, 1994).               
\bibitem{ola92}  Z. Olami, H. J. S. Feder and K. Christensen, 
                 Phys. Rev. Lett. {\bf 68}, 1244 (1992).
\bibitem{chr92}  K. Christensen and Z. Olami, 
                 Phys. Rev. A {\bf 46}, 1829 (1992).
\bibitem{bak89}  P. Bak, K. Chen, M. Creutz, 
                 Nature {\bf 342}, 789 (1989).
\bibitem{sne93}  K. Sneppen, and P. Bak, 
                 Phys. Rev. Lett. {\bf 71}, 4083 (1993).
\bibitem{nag92}  K. Nagel and E. Raschke, 
                 Physica A {\bf 182}, 519 (1992).
\bibitem{tak92}  H. Takayasu and H. Inaoka, 
                 Phys. Rev. Lett. {\bf 68}, 966 (1992).
\bibitem{dro92}  B. Drossel and F. Schwabl,  
                 Phys. Rev. Lett. {\bf 69}, 1629 (1992).
\bibitem{dro93a} B. Drossel and F. Schwabl, 
                 Physica A {\bf 199}, 183 (1993).
\bibitem{tys88}  J. J. Tyson and J. P. Keener,
                 Physica D {\bf 32}, 327 (1988). 
\bibitem{mer92}  E. Meron, 
                 Phys. Rep. {\bf 218}, 1 (1992).
\bibitem{bak90}  P. Bak, K. Chen and C. Tang,  
                 Phys. Lett. A {\bf 147}, 297 (1990).
\bibitem{gra91}  P. Grassberger and H. Kantz,  
                 J. Stat. Phys. {\bf 63}, 685 (1991).
\bibitem{mos92}  W. Mo\ss ner, B. Drossel, and F. Schwabl,  
                 Physica A {\bf 190}, 205 (1992).
\bibitem{gre78}  J. M. Greenberg and S. P. Hastings, 
                 Siam. J. Appl. Math. {\bf 34}, 515 (1978).
\bibitem{alb95}  E. V. Albano,
                 Physica A {\bf 216}, 213 (1995). 
\bibitem{joh94}  A. Johansen, 
                 Physica D {\bf 78}, 186 (1994); 
                 J. Theor. Biol. I {\bf 178}, 45 (1996).
\bibitem{hen89}  C. L. Henley, 
                 Bull. Am. Phys. Soc. {\bf 34}, 838 (1989).
\bibitem{hen93}  C. L. Henley, 
                 Phys. Rev. Lett. {\bf 71}, 2741 (1993).
\bibitem{cla94}  S. Clar, B. Drossel, and F. Schwabl,
                 Phys. Rev. E {\bf 50}, 1009 (1994). 
\bibitem{dro93}  B. Drossel, S. Clar, and F. Schwabl, 
                 Phys. Rev. Lett. {\bf 71}, 3739 (1993).
\bibitem{gra93}  P. Grassberger, 
                 J. Phys. A {\bf 26}, 2081 (1993).
\bibitem{chr93}  K. Christensen, H. Flyvberg, and Z. Olami, 
                 Phys. Rev. Lett. {\bf 71}, 2737 (1993).
\bibitem{dro94a} B. Drossel, S. Clar, and F. Schwabl,
                 Phys. Rev. E {\bf 50}, R2399 (1994).
\bibitem{dro94}  B. Drossel, 
                 Ph.D. Thesis, TU M\"unchen 1994.
\bibitem{dro94c} B. Drossel and F. Schwabl, 
                 Physica A {\bf 204}, 212 (1994).
\bibitem{sta92}  D. Stauffer and A. Aharony, 
                 {\em Introduction to Percolation Theory}, 
                 (Taylor and Francis, London, 1992).
\bibitem{pac93}  M. Paczuski and P. Bak,
                 Phys. Rev. E {\bf 48}, R3214 (1993).
\bibitem{dro94b} B. Drossel, S. Clar, and F. Schwabl,
                 Z. Naturforsch. {\bf 49}, 856 (1994).
\bibitem{dro95}  B. Drossel,
                 Phys. Rev. Lett. {\bf 76}, 936 (1996).
\bibitem{lor95}  V. Loreto, L. Pietronero, A. Vespignani, and S. Zapperi,
                 Phys. Rev. Lett. {\bf 75}, 465 (1995).
\bibitem{pat94}  H. Patzlaff and S. Trimper, 
                 Phys. Lett. A {\bf 189}, 187 (1994).
\bibitem{bot95}  S. Bottani and B. Delamotte,
                 to be published.
\bibitem{cla95}  S. Clar, B. Drossel, and F. Schwabl, 
                 Phys. Rev. Lett. {\bf 75}, 2722 (1995).
\bibitem{cla96}  S. Clar, K. Schenk, and F. Schwabl, 
                 to be published.
\bibitem{che90}  K. Chen, P. Bak, and M. Jensen,
                 Phys. Lett. A {\bf 149}, 207 (1990).
\bibitem{soc93}  J. E. S. Socolar, G. Grinstein, and C. Jayaprakash,
                 Phys. Rev. E {\bf 47}, 2366 (1993).
\bibitem{cha95}  T. C. Chan, H. F. Chau, and K. S. Cheng,
                 Phys. Rev. E {\bf 51}, 3045 (1995).
\bibitem{che90a} X. Che and H. Suhl,
                 Phys. Rev. Lett. {\bf 64}, 1670 (1990).
\bibitem{bak92} P. Bak and H. Flyvbjerg,
                Phys. Rev. A{\bf 45}, 2192 (1992).
\bibitem{ber95} A. T. Bernardes and J. G. Moreira,
                J. Phys. I France {\bf 5}, 1135 (1995).
\bibitem{odd93} L. Oddershede, P. Dimon and J. Bohr,
                Phys. Rev. Lett. {\bf 71}, 3107 (1993).
\bibitem{tan93} C. Tang,
                Physica A{\bf 194}, 315 (1993).
\bibitem{pac95}  M. Paczuski, S. Maslov, and P. Bak, 
                 to be published in Phys. Rev. E (1995).
\bibitem{dha90a} D. Dhar, 
                 Phys. Rev. Lett. {\bf 64}, 1613 (1990).
\bibitem{jan93}  I. M. Janosi and J. Kertesz,
                 Physica A {\bf 200}, 179 (1993).
\bibitem{gra94}  P. Grassberger, 
                 Phys. Rev. E {\bf 49}, 2436 (1994).
\bibitem{mid95}  A. A. Middleton and C. Tang, 
                 Phys. Rev. Lett. {\bf 74}, 742 (1995).
\bibitem{car89}  J. M. Carlson and J. S. Langer,
                 Phys. Rev. Lett. {\bf 62}, 2632 (1989);
                 Phys. Rev. A {\bf 40}, 6470 (1989).
\bibitem{hav91}  S. Havlin, A. L. Barabasi, S. V. Buldyrev, C. K. Peng,
                 M. Schwartz, H. E. Stanley and T. Vicsek,
                 in Proc. Granada Conf., Spain (Oct 7 - 11, 1991),
                 ed. by J. M. Garcia-Ruiz, E. Louis, L. Sander and P. Meakin 
                 (New York, Plenum, 1992).
\bibitem{sne92}  K. Sneppen, Phys. Rev. Lett. {\bf 69}, 3539 (1992).
\bibitem{zai92}  T. Zaitsev, Physica A{\bf 189}, 411 (1992). 
\end{references}
\end{document}